\documentclass[12pt]{article}
\usepackage[authoryear]{natbib}
\usepackage{cite}
\usepackage{setspace}
\usepackage[table]{xcolor}
\usepackage{tabularx,colortbl}
\usepackage{float}
\usepackage{caption}
\usepackage{booktabs,multirow,array}
\usepackage{chngpage}
\usepackage{subcaption}
\usepackage{csquotes}
\usepackage{array}
\newcolumntype{P}[1]{>{\centering\arraybackslash}p{#1}}
\usepackage{indentfirst}
\usepackage{graphicx, amsmath, bm, rotating, pdflscape, xcolor, longtable}
\usepackage{color}
\usepackage{amssymb}
\usepackage{dcolumn} % this is for package the generate latex tables from R output
\usepackage{hyperref}
\usepackage[margin=1in]{geometry}
\usepackage{siunitx}
\usepackage{etoolbox}
\usepackage[margin=20pt,font=footnotesize,labelfont=bf, textfont=bf,labelsep=endash
]{caption}
\usepackage{endnotes} 
\usepackage{authblk}
\usepackage{setspace}
\title{Explaining the Decline of Child Mortality in 44 Developing Countries: \\A Bayesian Extension of Oaxaca Decomposition Methods}

\author[1,2]{Antonio Pedro Ramos \thanks{Contact Author: tomramos@ucla.edu}}
\author[3]{Martiniano Jose Flores}
\author[1]{Leiwen Gao}
\author[4]{Patrick Heuveline}
\author[1]{Robert E. Weiss}
\affil[1]{Department of Biostatistics, Fielding School of Public Health, UCLA, Los Angeles, California, USA}
\affil[2]{ California Center for Population Research, UCLA, Los Angeles, California, USA}
\affil[3]{Department of General Internal Medicine, UCLA David Geffen School of
Medicine, Los Angeles, CA, 90095}
\affil[4]{Department of Sociology and Califonia Center for Population Research, Los Angeles, CA, 90095}

\begin{document}
\maketitle
\thispagestyle{empty}

\abstract{We investigate the decline of infant mortality in 42 low and middle income countries (LMIC) using detailed micro data from 84 Demographic and Health Surveys. We estimate infant mortality risk for each infant in our data and develop a novel extension of Oaxaca decomposition to understand the sources of these changes. We find that the decline in infant mortality is due to a declining propensity for parents with given characteristics to experience the death of an infant rather than due to changes in the distributions of these characteristics over time. Our results suggest that technical progress and policy health interventions in the form of public goods are the main drivers of the the recent decline in infant mortality in LMIC.}

\newpage

\section{Introduction}
\doublespacing

The United Nations Sustainable Development Goals (SDGs) for 2030 call for reduction of early-life mortality (ELM). In particular, it calls for all countries to reduce their neonatal (under 28 days) mortality to no more than 12 deaths per 1,000 live births and their under-5 mortality to no more than 25 deaths per 1,000 live births \citep{UN:2016}. Despite a 44\% reduction in child mortality globally from 2000 to 2015, there were still an estimated 5.9 million child (under-five) deaths in 2015 with a global child mortality rate of 43 deaths per 1,000 live births (\url{https://www.who.int/news-room/fact-sheets/detail/children-reducing-mortality}). Similarly, the neonatal mortality rate declined from 31 deaths per 1,000 live births in 2000 to 19 deaths per 1,000 live births in 2015, still well above SDG 3. Progress toward the SDGs has varied widely from country to country \citep{rajaratnam:2010}. Many countries, particularly in sub-Saharan Africa and Southeast Asia, still have high infant mortality rates, sometimes as high as 84 deaths per 1,000 live births.  To assess whether declines in Early-Life Mortality (ELM) can be expected to continue requires a better understanding of the determinants of these declines. 

Infant and child survival is known to be associated with several parental characteristics, e.g., whether a child’s mother has completed her primary education \citep{desai:1998,Andriano:2019} \citep{kamal:2012}. Were recent declines to be driven by increases in the proportion of mothers having completed primary school, for instance, future declines might depend on whether future educational gains can be expected or whether this proportion is now approaching 100\%. Similarly, ELM has been shown to be associated with younger maternal age at birth \citep{finlay:2011} and rural parental residence \citep{van:2009} \citep{sastry:1997}. Using national level data \citet{bishai:2016} find that most of the decline in child mortality was due to the expansion --- change in the distribution  --- of a broad array of child health determinants. Similarly, \citet{van:2009} used micro data from Sub-Saharan Africa and found that most of the gap in infant mortality between rural and urban populations can be explained by rural household disadvantage in the distribution of risk factors relative to urban households. A similar conclusion was reached by \citet{Saika:2013} who found that most of the rural-urban gap in child mortality in India can be explained by rural households' disadvantage in the distribution of factors. 

To our knowledge, no study to date has undertaken a systematic investigation of the role of distributional changes in parental, infant and geographic characteristics in change in ELM across Low and Middle Income Countries (LMIC) using micro data. Micro data allow for a more fine grained understanding of these phenomena than aggregate data because we can explore the effects of individual, parental and geographical characteristics on births directly unlike when working with aggregated data. Micro data also allow researchers to use random effects models that take into account unknown characteristics at different levels (e.g. individual, parental, and geographic). 

The shortage of studies using micro data may be due to the methodological challenges of using micro data, for example the need for non-linear random effects models to estimate mortality risk. Decomposition methods have been developed to separate the contribution of differences in distribution (typically referred to as populations’ “endowment”) and differences in relationship (typically referred to as “propensities”). Kitagawa (1955) introduced a ``categorical” form of decomposition for the difference between two aggregate rates using population distribution across groups (categories). Oaxaca (1973) introduced a “statistical” form of decomposition using linear regression models. A number of methodological developments have built on these two seminal contributions, but as discussed in the next section, none is fully adapted to take advantage of the wealth of micro data on parental, birth and geographic characteristics that are available in demographic and health surveys and that can be used to estimate mortality risk at the individual level using non-linear random effects models with geographic location effects. 

In this paper, we present new methods to undertake such a decomposition using predictions of infant mortality risk from a non-linear Bayesian hierarchical probit random effects model. We then apply these methods to data from 42 LMIC. We include countries with a variety of levels of infant mortality and estimate mortality risk using observable parental, individual and geographic characteristics available from Demographic and Health Surveys (DHS). 

We show that most of the decline in ELM mortality in LMIC in recent decades has been due to changes in the effects of the infant, parental and geographic characteristics (regressions coefficients) rather than changes in the distribution of these characteristics. Although large uncertainty in our estimates make it difficult to assess which characteristics mattered the most, our overall results suggest that the decline in child mortality is associated with general improvements in health, beyond the effects of the individual, parental, and geographic characteristics included in our analysis. Our results contrast with a large body of the previous literature which associate mortality declines with improvements in the distributions of risk factors. Instead, our findings suggest that most of the decline in Infant Mortality in recent decades in LMIC has been driven by technical progress. 

Our paper is organized as follows. Section \ref{data_and_methods} introduces the data used in our analyses and describes the statistical model used to estimate mortality risk. Section \ref{decom} introduces our new decomposition methods. Section \ref{results} reports on our empirical findings. Section \ref{discussion} discusses the methodological and policy implications of our findings. 

\section{Data and Statistical Model}
\label{data_and_methods}

\subsection{Data}

To investigate explanations for the decline in mortality under 1 year old, we assembled data from two waves of the Demographic and Health Surveys (DHS) for 42 countries. The two waves are between 10 and 20 years apart. For each survey, we include births from mothers aged 15 - 45 years old. We analyze births that occurred between one and five years before each survey to minimize censoring issues. Table 1 shows the survey year, sample size and empirical mortality rate in each wave for all 42 countries. We observe declines in mortality in all countries except Cameroon, which stayed the same from 1991 -- 2011.

To estimate the individual mortality risk in each survey within each country, we use a probit model including covariates that have been shown to have an association with child mortality, are available in all of the surveys, and are comparable across countries and surveys. We include maternal age in years, mother's education in years, gender of the infant, birth order, place of residence (rural versus urban), and relative wealth. Relative wealth is a continuous variable defined as the percentage of households with wealth less than or equal to that of the household where the mother lived at the time of the survey. We include a random effect for survey sampling cluster as a proxy for geographic location.

cluster

\subsection{Bayesian Hierarchical Model for Infant Mortality}
\label{model}

For each country, let $k \in \left\lbrace 1, 2\right\rbrace$ index surveys within country where $K_1$ is the earlier survey and $K_2$ is the later survey,  $j \in \left\lbrace 1, \dots, n_k \right\rbrace$ index clusters within survey, and $i \in \left\lbrace 1, \dots, N_{jk} \right\rbrace$ index birth within survey and cluster, where $n_k$ is the number of clusters in survey $k$ and $N_{jk}$ is the number of births in cluster $j$ and survey $k$. We omit country indices. The total number of infants in each survey is $N_k = \sum_{j=1}^{n_k} N_{jk}$. Let $y_{ijk}$ be a binary indicator for whether or not child $i$ died under 1 year old in cluster $j$ and survey $k$, with $y_{ijk} = 1$ if child $i$ died and $y_{ijk} = 0$ otherwise. We model $y_{ijk}$ as a Bernoulli random variable with mortality probability $\pi_{ijk}$,
\begin{align}
y_{ijk}|\pi_{ijk} &\sim \operatorname{Bern}(\pi_{ijk}),\label{eq:model1}
\end{align}
and a probit model for $\pi_{ijk}$,
\begin{align}
\text{P}(y_{ijk} = 1 | \bm{\beta}_k, \gamma_{jk}) = \Phi\left(\bm{x}_{ijk}^T\bm{\beta}_{k} + \gamma_{jk}\right),\label{eq:model2}
\end{align}
where $\Phi(\cdot)$ is the standard normal Cumulative Density Function (CDF), $\bm{x}_{ijk}$ is the covariate vector which includes an intercept term, and $\gamma_{jk}$ is a cluster level random effect that is normally distributed with variance $\sigma_{k}^2$,
\begin{align*}
\gamma_{jk}|\sigma^{2}_k &\sim \text{N}(0, \sigma_{k}^2).
\end{align*}

For interpretability, we center the continuous variables (maternal age, maternal education, birth order, and wealth) by subtracting off their respective means in the poorest 20\% of households in the first survey. We treat female births from rural places of residence as reference levels, giving the intercept an interpretation of the logit probability of death for the average poor, average age, average education, average birth order, rural, and female baby in the survey. Finally, to account for potential nonlinear mortality trends we fit B-splines to maternal age, maternal education, birth order, and relative wealth score. Our exact model is  presented in detail in the Appendix. 

\section{Oaxaca Decomposition}
\label{decom}

Oaxaca decomposition methods take two populations under consideration and fit separate regression models to each population to estimate the average response for each population. The average  difference between populations is decomposed into two parts, one that is due to covariate distribution effects (the distribution of the X's for example the distribution of parental, birth and geographic characteristics) and another that is  due to differences in the covariate-response relationship (regression coefficients which are the effects of those characteristics on infant mortality). In multivariate models, one can further decompose overall effects into individual $X$ and $\beta$ effects. 

In its original formulation, Oaxaca decomposition methods make use of the fact that in linear models, the average \emph{estimated} response in a population is equal to the response at the average covariate value in the population. However, to take advantage of the parental, individual, and geographic information available in surveys, mortality risk needs to be estimated using nonlinear logistic regression models. In these models, because of their non-linearity, the average \emph{estimated} response is not equal to the response at the average covariate value. Overcoming this limitation, \citet{fairlie:2005} developed an Oaxaca decomposition for logit and probit models, assuming matched samples between two population and fitting a model to the pooled data. \citet{Bauer:2008} built upon this approach and extended the Oaxaca decomposition to a more general class of non-linear models. \citet{van:2009} extended Oaxaca decompositions to binary response models with random effects for geographic locations (community effects).  They still assume matched samples and treat random effects as regular regression coefficients for their decomposition procedure but they do not provide uncertainty estimates for their decomposition results. 

Despite all these efforts a number of important methodological gaps remain in the application of Oaxaca decomposition methods to the decline in ELM. While it is possible to further decompose the covariate effect into the effects of individual covariates, to our knowledge similar decompositions do not exist for the effects of individual coefficients. Moreover, existing methods require matched samples between the two populations under consideration. This is not generally reasonable. Populations are not by nature matched, as, for example, populations from two separate countries, rural versus urban populations, or wealthy versus poor populations. Another undesirable consequence of the assumption of matched samples is that when the samples are of unequal size, it is necessary to duplicate elements from the larger population or duplicate elements from the smaller population, which can bias results. 

A related issue arises when we have models with random effects. The method proposed by \citet{van:2009} has random effects for location, but this assumes that locations are the same in the two populations. While location effects can be of scientific interest, oftentimes we are interested in the unconditional effect of the covariates on infant mortality, which requires us to marginalize out the random effects.  Further, locations do not align between surveys in different countries or even across surveys within the same country but from different years. 

Most importantly, none of the previous work propagates uncertainty from the estimation of the mortality risk stage to the decomposition analysis stage. This is important because we would like to know which decomposition results are statistically significant.

To understand how the outcomes differ between two different surveys, a standard approach is to decompose the difference between the average responses from the surveys into two parts using the Oaxaca decomposition. One part is due to the differences in the distribution of covariates and the second is due to the differences in the regression coefficients between two surveys.

\subsection{Overall Oaxaca Decomposition}
Suppose for subject $i$ in survey $k$, $y_{ik}$ is a continuous outcome variable defined on the real line with $x_{ik}$ as a covariate vector for subject $i$ in survey $k$, and we model $y_{ik}$ with linear regression
\begin{align*}
y_{ik} = \bm{x}_{ik}'\bm{\beta}_{k} + \epsilon_{ik},
\end{align*}
where $\bm{\beta}_{k}$ is a vector of regression coefficients for survey $k$, $ k = 1, 2$, and $\epsilon_{ik} \sim \text{N}(0,\sigma_{k}^2)$ is a normal error term with variance $\sigma_{k}^2$. Letting $\hat{\bm{\beta}}_k$ be the least squares estimate for $\bm{\beta}_k$, the mean response in survey $k$ is $\bar{Y}_k = \frac{1}{N_k}\sum_{i=1}^{N_k}\bm{x}_{ik}' \hat{\bm{\beta}}_k = \bm{\bar{x}}_{k}'\hat{\bm{\beta}}_{k}$, where $\bar{x}_k=N_{k}^{-1}\sum_{i \in k}^{N_{k}X_{ik}}$. The Oaxaca decomposition for the difference in means between the two surveys is
\begin{align}
\bar{Y}_{1} - \bar{Y}_{2} &= \bm{\bar{x}}_{1}'\hat{\bm{\beta}}_{1} - \bm{\bar{x}}_{2}'\hat{\bm{\beta}}_{2}\\
&= \underbrace{(\bm{\bar{x}}_{1}'\hat{\bm{\beta}}_{1} -  \bm{\bar{x}}_{2}'\hat{\bm{\beta}}_{1})}_{ \text{X effect}}+ \underbrace{(\bm{\bar{x}}_{2}'\hat{\bm{\beta}}_{1} - \bm{\bar{x}}_{2}'\hat{\bm{\beta}}_{2})}_{\text{coefficient effect}}. \label{eq:decomp1}
\end{align}
The first term in \eqref{eq:decomp1} represents the difference due to changes in the mean of the covariates, and the second term represents the difference due to changes in the regression coefficients.

A similar decomposition exists for a general nonlinear model $\text{E}\left[y_{ik} | \bm{\beta}_k\right] = F(\bm{x}_{ik}^T\bm{\beta}_k)$, where $F\left(\cdot\right)$ is a link function,
\begin{align}
\text{E}[Y_{i1}\ | \bm{\beta}_{k}\sigma_{1}^{2}]- \text{E}[Y_{i2}\ | \bm{\beta}_{k}\sigma_{2}^{2}]&= \sum_{i = 1}^{N_{1}}\frac{F\left(\bm{x}_{i1}^T\bm{\beta}_{1}\right)}{N_{1}} - \sum_{i = 1}^{N_{2}}\frac{F\left(\bm{x}_{i2}^T\bm{\beta}_{2}\right)}{N_{2}}\label{eq:decomp2}\\
&=  \underbrace{\left[\sum_{i = 1}^{N_{1}}\frac{F\left(\bm{x}_{i1}^T\bm{\beta}_{1}\right)}{N_{1}} - \sum_{i = 1}^{N_{2}}\frac{F\left(\bm{x}_{i2}^T\bm{\beta}_{1}\right)}{N_{2}}\right]}_{\text{X effext}} + \underbrace{\left[\sum_{i = 1}^{N_{2}}\frac{F\left(\bm{x}_{i2}^T\bm{\beta}_{1}\right)}{N_{2}} - \sum_{i = 1}^{N_{2}}\frac{F\left(\bm{x}_{i2}^T\bm{\beta}_{2}\right)}{N_{2}}\right]}_{\text{coefficient effect}}, \label{eq:decomp3}
\end{align}
As \citet{fairlie:2005} notes, for a probit model, taking $F(\cdot)$ to be the cumulative distribution function (CDF) of a standard normal random variable, the left hand side of \eqref{eq:decomp2} is only approximately equal to the right hand side, but the approximation is generally very close.

\subsubsection{Overall Decomposition for the Probit Model with Random Effects}

For each country, we want to decompose the difference in the estimated mortality rates between surveys $1$ and $2$ into covariate effects and coefficient effects. However, model \eqref{eq:model1}-\eqref{eq:model2} contains cluster random effects $\gamma_{jk}$. The numbers and alignments of clusters are
 not generally comparable between surveys. Therefore, we base our decompositions on the \emph{marginal} model by integrating out the random effects. Let $\tilde{\pi}_{ijk}$ be the probability of death in the marginal model. Then
\begin{align}
\tilde{\pi}_{ijk} &=\text{E}\left[\text{E}[y_{ijk} | \bm{\beta}_k, \gamma_{jk}]\right]\label{eq:cond_expectation}\\
&= \Phi(\bm{x}_{ijk}^{T}\tilde{\bm{\beta}}_{k})\label{eq:beta_tilde},
\end{align}
where $\tilde{\bm{\beta}}_{k} = \bm{\beta}_{k}(1 + \sigma_{k}^{2})^{1/2}$ and \eqref{eq:cond_expectation} follows from iterated expectation  and the proof of equation \eqref{eq:beta_tilde} is provided in the appendix. Thus, the marginal model for the probability that $y_{ijk} = 1$ is still a probit model with the regression coefficients multiplied by a correction factor of $\left(1+\sigma_{k}^2\right)^{1/2}$.

Substituting \eqref{eq:beta_tilde} for $F(\cdot)$ in \eqref{eq:decomp3}, the decomposition becomes
\begin{align}
\text{E}\left[\bar{y}_{1} - \bar{y}_{2}\right]  &\approx \underbrace{\left[ \sum_{j = 1}^{n_{1}}\sum_{i = 1}^{N_{j1}}\frac{\Phi\left(\bm{x}_{ij1}^T\tilde{\bm{\beta}}_{1}\right)}{N_{1}} -  \sum_{j = 1}^{n_{2}}\sum_{i = 1}^{N_{j2}}\frac{\Phi\left(\bm{x}_{ij2}^T\tilde{\bm{\beta}}_{1}\right)}{N_{2}}\right]}_{\text{X effect}} + \underbrace{\left[\sum_{j = 1}^{n_{2}}\sum_{i = 1}^{N_{j2}}\frac{\Phi\left(\bm{x}_{ij2}^T\tilde{\bm{\beta}}_{1}\right)}{N_{2}} -  \sum_{j = 1}^{n_{2}}\sum_{i = 1}^{N_{j2}}\frac{\Phi\left(\bm{x}_{ij2}^T\tilde{\bm{\beta}}_{2}\right)}{N_{2}}\right]}_{\text{beta effect}} \label{eq:decomp9}
\end{align}
	
\subsubsection{Coefficient By Coefficient Decomposition for the Probit model}
Once we have the overall decomposition, it can be of interest to explore the effect of individual regression coefficients on the overall decomposition. Suppose there are two covariates in the model and an intercept term. Then $\bm{x}_{ijk} = (1, x_{ijk1}, x_{ijk2})^{T}$ and  $\tilde{\bm{\beta}}_{k} = (\tilde{\beta}_{k0}, \tilde{\beta}_{k1}, \tilde{\beta}_{k2})^{T}$. The overall beta effect can be decomposed sequentially into the  contributions of each coefficient or set of coefficients as follows. The overall beta effect is	
\begin{align}
 &\left[\sum_{j = 1}^{n_{2}}\sum_{i = 1}^{N_{j2}}\frac{\Phi\left(\bm{x}_{ij2}^T\tilde{\bm{\beta}}_{1}\right)}{N_{j2}} - \sum_{j = 1}^{n_{2}}\sum_{i = 1}^{N_{j2}}\frac{\Phi\left(\bm{x}_{ij2}^T\tilde{\bm{\beta}}_{2}\right)}{N_{j2}}\right] \label{eq:overall}\\
&= \underbrace{\left[\sum_{j = 1}^{n_{2}}\sum_{i = 1}^{N_{j2}}\frac{\Phi\left(\tilde{\beta}_{10} + x_{ij21}\tilde{\beta}_{11} + x_{ij22}\tilde{\beta}_{12}\right)}{N_{j2}} - \sum_{j = 1}^{n_{2}}\sum_{i = 1}^{N_{j2}}\frac{\Phi\left(\tilde{\beta}_{20} + x_{ij21}\tilde{\beta}_{11} + x_{ij22}\tilde{\beta}_{12}\right)}{N_{j2}}\right]}_ {\text{the effect of intercept}} \label{eq:beta0}\\
& +\underbrace{\left[\sum_{j = 1}^{n_{2}}\sum_{i = 1}^{N_{j2}}\frac{\Phi\left(\tilde{\beta}_{20} + x_{ij21}\tilde{\beta}_{11} + x_{ij22}\tilde{\beta}_{12}\right)}{N_{j2}} - \sum_{j = 1}^{n_{2}}\sum_{i = 1}^{N_{j2}}\frac{\Phi\left(\tilde{\beta}_{20} + x_{ij21}\tilde{\beta}_{21} + x_{ij22}\tilde{\beta}_{12}\right)}{N_{j2}}\right]}_ {\text{the effect of first coefficent}}  \label{eq:beta1}\\
& +\underbrace{\left[\sum_{j = 1}^{n_{2}}\sum_{i = 1}^{N_{j2}}\frac{\Phi\left(\tilde{\beta}_{20} + x_{ij21}\tilde{\beta}_{21} + x_{ij22}\tilde{\beta}_{12}\right)}{N_{j2}} - \sum_{j = 1}^{n_{2}}\sum_{i = 1}^{N_{j2}}\frac{\Phi\left(\tilde{\beta}_{20} + x_{ij21}\tilde{\beta}_{21} + x_{ij22}\tilde{\beta}_{22}\right)}{N_{j2}}\right]}_{\text{the effect of second coefficent}} \label{eq:beta2}.
\end{align}

\noindent In equations \eqref{eq:overall} -- \eqref{eq:beta2}, \eqref{eq:beta0} is the portion of the decomposition due to differences in the intercept where $\beta_{j10}$ in the first term is replaced by $\beta_{j20}$ in the second term. Then we keep the replacement of intercept from the first term of \eqref{eq:beta1}, and only replace $\beta_{j11}$ by  $\beta_{j12}$ in the second term. Similarly, the replacement of intercept and first coefficient is kept in the first term of \eqref{eq:beta2}, and $\beta_{j22}$ substitutes for $\beta_{j21}$ in the second term. As a result, \eqref{eq:beta1} and \eqref{eq:beta2} are the portions of the decomposition that are due to changes of the regression coefficients individually. 

There are 7 covariates in our model, so to do the decomposition, we expand the three equations in \eqref{eq:beta0}-\eqref{eq:beta2} to eight, substituting one coefficient (or set of coefficients for categorical variables) in each equation. In general, the order in which we decompose the overall beta effect into the individual beta effects will matter.

\subsubsection{Accounting for Uncertainty in the Decomposition Results}
We estimate mortality risk for subjects with the model \eqref{eq:model1} -- \eqref{eq:model2} using Markov Chain Monte Carlo (MCMC). For each iteration $\ell = 1, \ldots, L$ of the sampler, we have random draws $\bm{\beta}_1^{(\ell)}$ and $\bm{\beta}_2^{(\ell)}$ from the posterior distribution of $\bm{\beta}_1$ and $\bm{\beta}_2$. Substitution of the estimates in place of the true parameters in \eqref{eq:decomp3} and \eqref{eq:beta0} -- \eqref{eq:beta2} gives us a posterior distribution for the components of the overall decomposition and for the components of the individual coefficient by coefficient decompositions, providing a straightforward way to get point estimates and measures of uncertainty.

\section{Results}
\label{results}
	
We fit the Bayesian model to predict infant mortality using Markov Chain Monte Carlo methods the \texttt{MCMCglmm} package in \texttt{R} \citep{Hadfield:2010,R}, which allows us to get posterior samples of infant mortality $\pi_{ijk}$. We ran the model until we obtained 1250 approximately independent posterior samples. We assessed convergence using trace plots and autocorrelation using ACF plots and convergency seems acceptable. 
	
\subsection{Decline of Child Mortality Between Two Surveys}
Table 2 gives posterior summaries of the mortality distributions for all countries and surveys. Countries are ordered by estimated yearly decline in infant mortality. Mortality rates are presented as deaths per 1,000 births. The `Years Between' column gives the number of years between the two surveys. The 'S1' and 'S2' columns indicate the estimated number of deaths per 1000 births for surveys 1 and 2, respectively. The `Diff' column is the difference in the number of deaths per thousand births between survey 1 and 2, and the `Yearly' column is `Difference' divided by `Years between', giving an estimate of how fast the mortality rate has decreased annually. The `95\% CI' columns denote 95\% posterior intervals. 

Comparing the model output with the raw data, we can see that the estimated average mortality rates presented in Table 2 are consistent with the raw mortality rates from Table 1. The general trend is that there are substantive and significant decreases in the average mortality over time for all countries except Cameroon, Zimbabwe, and Colombia. However, there is heterogeneity between countries in the yearly decline in infant mortality, from a decline of more than 5 deaths per 1000 births per year in Cambodia and Niger to declines of less than one death per 1000 births in Indonesia, Pakistan, Philippines, and Jordan.    
	
\subsection{Decomposition Results}

Overall covariate and coefficient effects as well as uncertainty estimates are presented in Table 3. Countries are ordered as  in Table 2. The `Effects' columns denote the actual contribution of the covariate and coefficient effects to the overall annualized decline, and the `\%' column denotes the percent of the overall decline that is due to the covariate and coefficient effects, with significant effects in boldface. For example the first row of Table 3 says that for Cambodia from survey 1 to survey 2 the mortality rate decreased by about 5\% ($X$ effect of 1\% + beta effect of 4.4\%). The percent reduction due to changes in the covariate distributions was  18\% and the reduction due to changes in the covariate response accounted for 82\%. 
	
The individual percentages can be more than 100\% when for example a beneficial covariate effect is more than offset by a deleterious coefficient effect or vice versa. For example, in Cote d'Ivoire, covariate effects led to an \emph{increase} in the mortality rate of 4 deaths per thousand births, but this was offset by the coefficient effect, which led to a \emph{decrease} in the mortality rate of 34 deaths per thousand births. Similar trends occur in Tanzania, Senegal, Burkina Faso, Gabon, and the Dominican Republic. For the remaining 35 countries excluding Jordan, Armenia, Ghana and Haiti, the decline in mortality risk is mostly explained by the coefficient effects. For countries with positive overall covariate effects, Kenya has the largest coefficient effect ($99\%$, 95\% CI $48\%, 161\%)$. 
	
Table 4 presents the results for the coefficient by coefficient decompositions. Countries are ordered as in Tables 2 and 3. Uncertainty is large for these decompositions and the vast majority are not significant. The intercept is significant for six out of 39 countries: Mali (155\%), Benin (165\%), Comoros (117\%), Tanzania (155\%), Uganda (118\%), and Kenya (116\%). A discussion of the statistical significance in the beta-by-beta decomposition can be found in the appendix A3.
		
\section{Discussion and Conclusions}
\label{discussion}

We use detailed information from DHS surveys on infants from a large number of LMIC's to understand drivers of infant mortality decline in recent decades. Understanding the mechanisms for the decline in child mortality is important not only to explain one of the defining events of our time but also to design effective policies that can help countries meet the SDG's.

Our main results show that for most countries the decline in infant mortality has been associated with actual declines in the propensities of parents with given characteristics to experience an infant death rather than changes in the distribution of these characteristics. This is in contrast to the vast majority of the previous work on the topic find an association between differences in average mortality risk between populations with differences in the distributions of infant, parental, and geographical characteristics over time, not to changes in the effects of these characteristics over time \citep{van:2009,bishai:2016,Demombynes:2016}.

We also find substantial between--country variation in the relative contribution of these declining propensities to the overall decline in infant mortality. For example, the effects of declining propensities on infant mortality amount to as little as 56\% of the actual decline as seen in Peru, and as much as 110\% of the actual decline in Cote d’Ivoire. The latter implies that this contribution is counterbalanced by distributional changes in parental characteristics that contributed to increased infant mortality in Cote d’Ivoire.  

Because most of the decline was due to changes in the effects of individual, parental and geographic characteristics, as well as the intercept --- the betas from the statistical model ---, we tried to isolate the contribution of each one of them to the overall decline. Although we were unable to find statistically significant effects for most of these characteristics, some patterns are clear. In general, the changes in the intercept tend to have the largest effect on mortality declines. The intercept represents the part of the model that captures the effects of the characteristics not included in the model. 

We interpret our findings as suggesting that most of the decline over time in infant mortality was due to technical progress and public health interventions in the form of public goods. The relative importance of the variables not included in the model suggests that general improvements in health in the form of public goods have increased the survival rates for many births. Examples of such public goods abound: vaccination campaigns, increased access to clean water and sanitation, health education programs, malaria control programs, and family planning \citep{Jamison:2016,Cutler:2006b,soares:2007}. Although these interventions are usually targeted to specific diseases or risk factors their benefits are often not confined to specific subgroups.  

On the other hand, that fact that births that were born under the same conditions experience lower mortality suggest technical progress. Parents living in the same location and with the same characteristics are now able to better take care of their offspring, perhaps because they are more knowledgeable about health and health care. 

If our interpretation is correct, the global health community committed to the mortality reductions for the SDG should direct their efforts toward policies that can benefit large groups of births, for example, family planning, educational initiatives, and vaccination campaigns. This has the added benefit of potentially leading to spillover effects over multiple communities. New medical technologies should be introduced as public goods. Moreover, empowering parents with knowledge and information can potentially further reduce mortality. That said, and given the between country heterogeneity, countries should rank their priorities according to their local conditions.

\newpage
%\bibliographystyle{chicagomjf}
%\bibliography{decomposition}

\newpage
	
	\begin{table}[H]
		\centering
		Table 1. Summary of dataset for the two surveys in 42 countries. The last column shows the order of the country in Table 2.
		\vspace{0.1cm}
		\singlespacing
		\footnotesize
		\begin{tabular}{ccccccccc}
			\toprule
			Country  & \multicolumn{3}{c}{Survey 1} & \multicolumn{3}{c}{Survey 2}\\
			& Year& Sample size & Deaths / 1000& Year& Sample size & Deaths / 1000&Order\\
			\midrule
			Armenia & 2000  & 1453  & 40    & 2010  & 1077  & 14    & 13 \\
			Bangladesh & 2000  & 5323  & 69    & 2014  & 7733  & 41    & 20 \\
			Benin & 1996  & 3968  & 106   & 2012  & 10361 & 48    & 7 \\
			Bolivia & 1998  & 5755  & 66    & 2008  & 6821  & 45    & 19 \\
			Burkina Faso & 1993  & 4363  & 103   & 2010  & 11701 & 75    & 24 \\
			Cambodia & 2000  & 6929  & 103   & 2014  & 5590  & 29    & 1 \\
			Cameroon & 1991  & 2546  & 67    & 2011  & 8937  & 67    & 40 \\
			Chad  & 1997  & 5552  & 118   & 2015  & 13697 & 72    & 12 \\
			Colombia & 1990  & 2930  & 23    & 2005  & 10975 & 22    & 42 \\
			Comoros & 1996  & 1607  & 78    & 2012  & 2345  & 30    & 9 \\
			Cote dIvoire & 1999  & 1367  & 116   & 2012  & 5915  & 81    & 11 \\
			Dominican Republic & 1996  & 1470  & 44    & 2013  & 3623  & 31    & 34 \\
			Egypt & 1995  & 7402  & 71    & 2014  & 12043 & 24    & 15 \\
			Gabon & 2000  & 2525  & 61    & 2012  & 4552  & 40    & 26 \\
			Ghana & 1993  & 2303  & 75    & 2014  & 4464  & 51    & 35 \\
			Guatemala & 1999  & 3782  & 50    & 2015  & 9099  & 28    & 28 \\
			Guinea & 1999  & 4805  & 109   & 2012  & 5367  & 77    & 17 \\
			Haiti & 1994  & 2215  & 79    & 2012  & 5586  & 65    & 33 \\
			India & 1993  & 42701 & 73    & 2006  & 40833 & 52    & 29 \\
			Indonesia & 1997  & 14662 & 49    & 2012  & 14244 & 36    & 36 \\
			Jordan & 1990  & 6721  & 32    & 2012  & 8585  & 22    & 39 \\
			Kenya & 1993  & 4833  & 61    & 2014  & 16607 & 39    & 32 \\
			Kyrgyzstan & 1997  & 1616  & 59    & 2012  & 3197  & 30    & 21 \\
			Madagascar & 1997  & 4721  & 104   & 2009  & 9294  & 49    & 4 \\
			Malawi & 1992  & 3365  & 138   & 2015  & 11429 & 42    & 5 \\
			Mali  & 1996  & 7354  & 134   & 2012  & 6823  & 62    & 3 \\
			Morocco & 1992  & 4171  & 58    & 2003  & 4101  & 41    & 27 \\
			Mozambique & 1997  & 5614  & 120   & 2011  & 8265  & 67    & 8 \\
			Namibia & 1992  & 2948  & 66    & 2013  & 3812  & 46    & 31 \\
			Niger & 1998  & 6152  & 133   & 2012  & 9530  & 62    & 2 \\
			Nigeria & 1990  & 6005  & 103   & 2013  & 26019 & 73    & 30 \\
			Pakistan & 1991  & 4614  & 85    & 2012  & 7862  & 69    & 37 \\
			Peru  & 1992  & 6468  & 62    & 2012  & 18988 & 20    & 23 \\
			Philippines & 1993  & 7315  & 39    & 2013  & 5841  & 27    & 38 \\
			Rwanda & 1992  & 4454  & 86    & 2015  & 5579  & 31    & 14 \\
			Senegal & 1997  & 5477  & 79    & 2015  & 7645  & 38    & 22 \\
			Tanzania & 1999  & 3802  & 91    & 2015  & 6430  & 41    & 10 \\
			Togo  & 1998  & 5497  & 90    & 2014  & 4950  & 50    & 18 \\
			Turkey & 1993  & 2926  & 57    & 2004  & 3405  & 34    & 16 \\
			Uganda & 1995  & 5568  & 90    & 2011  & 6178  & 58    & 25 \\
			Zambia & 1996  & 4335  & 118   & 2013  & 9254  & 50    & 6 \\
			Zimbabwe & 1994  & 3316  & 59    & 2015  & 4893  & 55    & 41 \\
			\bottomrule
		\end{tabular}
	\end{table}

	\setlength\LTleft{-0.9cm}
	\begin{table}[H]
		\centering
		Table 2. Estimated mortality risk per 1000 births in two surveys for 42 countries. Countries are ordered from largest to smallest yearly decline in child mortality. \textit{Years between} is the time in years between the two surveys. \textit{S1} and \textit{S2} are the estimated mortality risk per 1000 births in survey 1 and survey 2, respectively. \textit{Diff per year} is the difference in mortality risk per 1000 births, \textit{S1} minus \textit{S2}, per year.
		\vspace{0.1cm}
		\singlespacing
		\scriptsize
		\sisetup{
			table-align-uncertainty=true,
			separate-uncertainty=true,
		}
		%% local redefinitions
		\renewrobustcmd{\bfseries}{\fontseries{b}\selectfont}
		\renewrobustcmd{\boldmath}{}
		\begin{tabular}{
				c
				c
				S[table-format = 1, detect-weight,mode=text]
				>{{[}} % Add square bracket before column
				S[table-format = -2,table-space-text-pre={[}]
				@{,\,} % Add comma and thin-space between the columns
				S[table-format = -1.1,table-space-text-post={]}]
				<{{]}} % Add square bracket after column
				S[table-format = 1, detect-weight,mode=text]
				>{{[}} % Add square bracket before column
				S[table-format = -2,table-space-text-pre={[}]
				@{,\,} % Add comma and thin-space between the columns
				S[table-format = -1.1,table-space-text-post={]}]
				<{{]}} % Add square bracket after column
				S[table-format = 1, detect-weight,mode=text]
				>{{[}} % Add square bracket before column
				S[table-format = -2,table-space-text-pre={[}]
				@{,\,} % Add comma and thin-space between the columns
				S[table-format = -1.1,table-space-text-post={]}]
				<{{]}} % Add square bracket after column
				S[table-format = 1, detect-weight,mode=text]
				>{{[}} % Add square bracket before column
				S[table-format = -1.1,table-space-text-pre={[}]
				@{,\,} % Add comma and thin-space between the columns
				S[table-format = -1.2,table-space-text-post={]}]
				<{{]}} % Add square bracket after column
			}
			\hline
			&Years&\multicolumn{6}{c}{Mean mortality probability (per 1000)}&\\
			Country & between  & {S1} & \multicolumn{2}{p{2cm}}{95\% CI} &{S2} & \multicolumn{2}{p{2cm}}{95\% CI} &{Diff} & \multicolumn{2}{p{2cm}}{95\% CI}&{Diff per year} & \multicolumn{2}{p{1cm}}{95\% CI} \cr
			\hline
			Cambodia & 14    & 106   & 96    & 115   & 31    & 25    & 37    & \bfseries75 & 63    & 87    & \bfseries5.4 & 4.5   & 5.8 \\
			Niger & 14    & 135   & 122   & 149   & 64    & 57    & 71    & \bfseries71 & 56    & 87    & \bfseries5.1 & 4.0   & 6.2 \\
			Mali  & 16    & 141   & 129   & 154   & 64    & 56    & 73    & \bfseries77 & 62    & 92    & \bfseries4.8 & 3.9   & 5.8 \\
			Madagascar & 12    & 107   & 93    & 121   & 51    & 45    & 57    & \bfseries56 & 41    & 72    & \bfseries4.7 & 3.4   & 6.0 \\
			Malawi & 23    & 144   & 126   & 161   & 43    & 39    & 48    & \bfseries100 & 82    & 118   & \bfseries4.3 & 3.6   & 5.1 \\
			Zambia & 17    & 123   & 110   & 137   & 52    & 46    & 58    & \bfseries71 & 57    & 86    & \bfseries4.2 & 3.4   & 5.1 \\
			Benin & 16    & 114   & 98    & 131   & 49    & 44    & 55    & \bfseries65 & 48    & 83    & \bfseries4.1 & 2.5   & 5.2 \\
			Mozambique & 14    & 125   & 113   & 139   & 70    & 62    & 77    & \bfseries56 & 41    & 70    & \bfseries4.0 & 2.9   & 5.0 \\
			Comoros & 16    & 90    & 71    & 115   & 35    & 26    & 46    & \bfseries56 & 33    & 81    & \bfseries3.5 & 2.1   & 5.1 \\
			Tanzania & 16    & 97    & 84    & 111   & 44    & 38    & 51    & \bfseries53 & 37    & 68    & \bfseries3.3 & 2.3   & 4.3 \\
			Cote dIvoire & 13    & 123   & 99    & 150   & 83    & 73    & 93    & \bfseries41 & 14    & 69    & \bfseries3.2 & 1.1   & 5.3 \\
			Chad  & 18    & 123   & 110   & 137   & 74    & 68    & 80    & \bfseries50 & 34    & 65    & \bfseries2.8 & 1.9   & 3.6 \\
			Armenia & 10    & 43    & 31    & 59    & 17    & 9     & 28    & \bfseries26 & 9     & 43    & \bfseries2.6 & 0.9   & 4.3 \\
			Rwanda & 23    & 93    & 80    & 107   & 34    & 28    & 40    & \bfseries59 & 44    & 74    & \bfseries2.6 & 1.9   & 3.2 \\
			Egypt & 19    & 72    & 65    & 79    & 25    & 22    & 28    & \bfseries48 & 39    & 56    & \bfseries2.5 & 2.1   & 2.9 \\
			Turkey & 11    & 61    & 50    & 72    & 34    & 26    & 42    & \bfseries27 & 14    & 41    & \bfseries2.5 & 1.3   & 3.7 \\
			Guinea & 13    & 113   & 100   & 127   & 82    & 72    & 94    & \bfseries31 & 14    & 49    &\bfseries2.4 & 1.1   & 3.8 \\
			Togo  & 16    & 92    & 80    & 105   & 55    & 46    & 64    & \bfseries37 & 22    & 53    & \bfseries2.3 & 1.4   & 3.3 \\
			Bolivia & 10    & 70    & 62    & 78    & 47    & 40    & 53    & \bfseries23 & 12    & 34    & \bfseries2.3 & 1.2   & 3.4 \\
			Bangladesh & 14    & 73    & 64    & 83    & 43    & 38    & 50    & \bfseries30 & 19    & 41    & \bfseries2.1 & 1.4   & 2.9 \\
			Kyrgyzstan & 15    & 66    & 49    & 85    & 34    & 26    & 45    & \bfseries32 & 13    & 52    & \bfseries2.1 & 0.9   & 3.5 \\
			Senegal & 18    & 82    & 72    & 93    & 44    & 37    & 52    & \bfseries38 & 26    & 51    & \bfseries2.1 & 1.4   & 2.8 \\
			Peru  & 20    & 63    & 56    & 70    & 20    & 18    & 23    & \bfseries42 & 34    & 50    & \bfseries2.1 & 1.7   & 2.5 \\
			Burkina Faso & 17    & 110   & 96    & 127   & 76    & 70    & 83    & \bfseries34 & 19    & 51    & \bfseries2.0 & 1.1   & 3.0 \\
			Uganda & 16    & 94    & 83    & 105   & 62    & 54    & 71    & \bfseries32 & 18    & 46    & \bfseries2.0 & 1.1   & 2.9 \\
			Gabon & 12    & 68    & 56    & 82    & 44    & 37    & 53    & \bfseries23 & 8     & 39    & \bfseries1.9 & 0.7   & 3.3 \\
			Morocco & 11    & 63    & 51    & 77    & 44    & 36    & 53    & \bfseries19 & 3     & 35    & \bfseries1.7& 0.3   & 3.2 \\
			Guatemala & 16    & 54    & 44    & 65    & 30    & 26    & 34    & \bfseries25 & 14    & 36    & \bfseries1.6 & 0.9   & 2.3 \\
			India & 13    & 72    & 70    & 76    & 52    & 49    & 55    & \bfseries20 & 17    & 24    & \bfseries1.5 & 1.3   & 1.8 \\
			Nigeria & 23    & 104   & 92    & 116   & 75    & 70    & 79    & \bfseries29 & 16    & 42    & \bfseries1.3 & 0.7   & 1.8 \\
			Namibia & 21    & 74    & 61    & 89    & 47    & 39    & 56    & \bfseries26 & 11    & 44    & \bfseries1.2 & 0.5   & 2.1 \\
			Kenya & 21    & 65    & 56    & 75    & 40    & 37    & 43    &\bfseries25 & 15    & 35    &\bfseries1.2 & 0.7   & 1.7 \\
			Haiti & 18    & 91    & 74    & 110   & 69    & 61    & 78    & \bfseries21 & 2     & 42    & \bfseries1.2 & 0.1   & 2.3 \\
			Dominican Republic & 17    & 52    & 38    & 68    & 33    & 27    & 40    & \bfseries19 & 4     & 36    & \bfseries1.1 & 0.2   & 2.1 \\
			Ghana & 21    & 78    & 65    & 94    & 56    & 47    & 66    & \bfseries22 & 5     & 41    & \bfseries1.0 & 0.2   & 2.0 \\
			Indonesia & 15    & 50    & 46    & 54    & 37    & 33    & 40    & \bfseries13 & 8     & 19    & \bfseries0.9 & 0.5   & 1.3 \\
			Pakistan & 21    & 90    & 79    & 101   & 72    & 64    & 79    & \bfseries18 & 5     & 32    & \bfseries0.9 & 0.2   & 1.5 \\
			Philippines & 20    & 41    & 36    & 47    & 29    & 24    & 35    & \bfseries12 & 4     & 20    & \bfseries0.6 & 0.2   & 1.0 \\
			Jordan & 22    & 35    & 29    & 41    & 23    & 19    & 27    & \bfseries12 & 4     & 19    & \bfseries0.5 & 0.2   & 0.9 \\
			Cameroon & 20    & 77    & 62    & 94    & 68    & 61    & 75    & 9     & -7    & 27    & 0.5   & -0.4  & 1.4 \\
			Zimbabwe & 21    & 65    & 53    & 78    & 58    & 50    & 67    & 7     & -9    & 21    & 0.3   & -0.4  & 1.0 \\
			Colombia & 15    & 26    & 20    & 34    & 23    & 20    & 26    & 3     & -4    & 11    & 0.2   & -0.3  & 0.7 \\
			\hline
		\end{tabular}
	\end{table}

	\begin{table}[H]
		\centering
		Table 3. Results of the decomposition in the decline of infant mortality risk per 1000 births per year in 42 countries. Countries are in the same order as Table 2. Bolded results are significant defined as zero not in the 95\% confidence interval.
		\vspace{0.1cm}
		\singlespacing
		\footnotesize
		\sisetup{
			table-align-uncertainty=true,
			separate-uncertainty=true,
		}
		%% local redefinitions
		\renewrobustcmd{\bfseries}{\fontseries{b}\selectfont}
		\renewrobustcmd{\boldmath}{}
		\begin{tabular}{
				c
				S[table-format = 1, detect-weight,mode=text]
				>{{[}} % Add square bracket before column
				S[table-format = -1.1,table-space-text-pre={[}]
				@{,\,} % Add comma and thin-space between the columns
				S[table-format = -1.2,table-space-text-post={]}]
				<{{]}} % Add square bracket after column
				S[table-format = 1, detect-weight,mode=text]
				>{{[}} % Add square bracket before column
				S[table-format = -2,table-space-text-pre={[}]
				@{,\,} % Add comma and thin-space between the columns
				S[table-format = -1.1,table-space-text-post={]}]
				<{{]}} % Add square bracket after column
				S[table-format = 1, detect-weight,mode=text]
				>{{[}} % Add square bracket before column
				S[table-format = -1.1,table-space-text-pre={[}]
				@{,\,} % Add comma and thin-space between the columns
				S[table-format = -1.2,table-space-text-post={]}]
				<{{]}} % Add square bracket after column
				S[table-format = 1, detect-weight,mode=text]
				>{{[}} % Add square bracket before column
				S[table-format = -3,table-space-text-pre={[}]
				@{,\,} % Add comma and thin-space between the columns
				S[table-format = -1.1,table-space-text-post={]}]
				<{{]}} % Add square bracket after column
			}
			\hline
			&\multicolumn{6}{c}{Overall X effects }&\multicolumn{6}{c}{Overall beta effects }\\
			Country    & {Effects} & \multicolumn{2}{p{2cm}}{95\% CI} &{$\%$} & \multicolumn{2}{p{2cm}}{95\% CI($\%$)} &{Effects} & \multicolumn{2}{p{2cm}}{95\% CI}&{$\%$} & \multicolumn{2}{p{2.4cm}}{95\% CI($\%$)} \cr
			\hline
            Cambodia & \bfseries1.0 & 0.4   & 1.5   & \bfseries18 & 8     & 29    & \bfseries4.4 & 3.5   & 5.3   & \bfseries82 & 65    & 100 \\
            Niger & \bfseries0.2 & 0.0   & 0.4   & \bfseries4 & 0     & 8     & \bfseries4.9 & 3.8   & 6.0   & \bfseries96 & 75    & 119 \\
            Mali  & 0.2   & -0.2  & 0.5   & 4     & -3    & 10    & \bfseries4.6 & 3.6   & 5.6   & \bfseries96 & 76    & 117 \\
            Madagascar & \bfseries0.8 & 0.5   & 1.2   & \bfseries18 & 11    & 25    & \bfseries3.8 & 2.6   & 5.1   & \bfseries82 & 55    & 108 \\
            Malawi & 0.0   & -0.6  & 0.5   & 0     & -13   & 12    & \bfseries4.4 & 3.4   & 5.3   & \bfseries100 & 78    & 122 \\
            Zambia & 0.2   & -0.2  & 0.6   & 5     & -5    & 14    & \bfseries4.0 & 3.1   & 4.9   & \bfseries95 & 74    & 118 \\
            Benin & \bfseries0.4 & 0.1   & 0.8   & \bfseries11 & 2     & 20    & \bfseries3.6 & 2.6   & 4.7   & \bfseries89 & 64    & 117 \\
            Mozambique & \bfseries0.6 & 0.1   & 1.1   & \bfseries15 & 3     & 28    & \bfseries3.4 & 2.3   & 4.5   & \bfseries85 & 57    & 114 \\
            Comoros & 0.5   & -0.1  & 1.1   & 16    & -2    & 31    & \bfseries2.9 & 1.5   & 4.6   & \bfseries85 & 42    & 131 \\
            Tanzania & -0.1  & -0.4  & 0.2   & -3    & -12   & 6     & \bfseries3.4 & 2.4   & 4.4   & \bfseries103 & 72    & 132 \\
            Cote dIvoire & -0.3  & -1.3  & 0.6   & -10   & -42   & 21    & \bfseries3.4 & 1.1   & 6.1   & \bfseries110 & 35    & 196 \\
            Chad  & 0.5   & -0.4  & 1.3   & 17    & -14   & 46    & \bfseries2.3 & 1.1   & 3.5   & \bfseries83 & 41    & 128 \\
            Armenia & \bfseries1.1 & 0.4   & 2.0   & \bfseries44 & 15    & 76    & 1.5   & -0.1  & 3.0   & 56    & -4    & 113 \\
            Rwanda & 0.3   & -0.1  & 0.7   & 11    & -5    & 26    & \bfseries2.3 & 1.6   & 3.0   &\bfseries89 & 61    & 119 \\
            Egypt & \bfseries0.9 & 0.4   & 1.3   & \bfseries35 & 16    & 51    & \bfseries1.6 & 1.1   & 2.2   & \bfseries65 & 43    & 88 \\
            Turkey & 0.2   & -0.2  & 0.5   & 8     & -9    & 22    & \bfseries2.2 & 1.0   & 3.6   & \bfseries93 & 42    & 148 \\
            Guinea & \bfseries0.3 & 0.1   & 0.6   & \bfseries13 & 3     & 24    & \bfseries2.1 & 0.8   & 3.4   & \bfseries87 & 33    & 144 \\
            Togo  &\bfseries0.4 & 0.0   & 0.7   & 17    & -1    & 32    & \bfseries1.9 & 1.0   & 2.9   & \bfseries84 & 43    &\bfseries127 \\
            Bolivia & 0.5   & -0.6  & 1.2   & 22    & -25   & 51    & \bfseries1.8 & 0.5   & 3.2   & \bfseries78 & 23    & 137 \\
            Bangladesh & \bfseries0.6 & 0.1   & 1.1   & \bfseries30 & 6     & 52    & \bfseries1.5 & 0.7   & 2.4   & \bfseries70 & 31    & 113 \\
            Kyrgyzstan & 0.2   & -0.7  & 1.0   & 9     & -34   & 46    & \bfseries1.9 & 0.5   & 3.7   & \bfseries91 & 23    & 174 \\
            Senegal & -0.1  & -0.4  & 0.3   & -4    & -20   & 12    & \bfseries2.2 & 1.4   & 3.0   & \bfseries104 & 66    & 142 \\
            Peru  & \bfseries0.9 & 0.6   & 1.2   & \bfseries44 & 29    & 58    & \bfseries1.2 & 0.8   & 1.6   &\bfseries56 & 38    & 77 \\
            Burkina Faso & -0.1  & -0.4  & 0.1   & -6    & -20   & 8     &\bfseries2.1 & 1.2   & 3.2   & \bfseries106 & 58    & 158 \\
            Uganda & \bfseries0.3 & 0.0   & 0.6   & \bfseries16 & 2     & 29    & \bfseries1.7 & 0.8   & 2.5   & \bfseries84 & 39    & 128 \\
            Gabon & -0.1  & -0.5  & 0.3   & -5    & -24   & 13    & \bfseries2.0 & 0.8   & 3.5   & \bfseries105 & 39    & 177 \\
            Morocco & 0.3   & -0.2  & 0.8   & 18    & -14   & 47    & \bfseries1.4 & 0.0   & 2.9   & \bfseries82 & 2     & 171 \\
            Guatemala & \bfseries0.6 & 0.1   & 1.0   & \bfseries37 & 10    & 62    & \bfseries1.0 & 0.3   & 1.7   & \bfseries63 & 18    & 113 \\
            India & \bfseries0.3 & 0.2   & 0.4   & \bfseries19 & 10    & 28    & \bfseries1.3 & 1.0   & 1.6   & \bfseries81 & 61    & 102 \\
            Nigeria & \bfseries0.3 & 0.1   & 0.6   & \bfseries27 & 7     & 44    & \bfseries0.9 & 0.3   & 1.5   & \bfseries74 & 28    & 120 \\
            Namibia & 0.3   & -0.3  & 0.8   & 26    & -21   & 68    & \bfseries0.9 & 0.0   & 1.9   & \bfseries74 & 4     & 151 \\
            Kenya & 0.0   & -0.5  & 0.4   & 1     & -43   & 36    & \bfseries1.2 & 0.6   & 1.9   &\bfseries99 & 48    & 161 \\
            Haiti & 0.2   & -0.5  & 0.9   & 17    & -44   & 73    & 1.0   & -0.2  & 2.3   & 84    & -18   & 197 \\
            Dominican Republic & -0.3  & -0.9  & 0.2   & -29   & -78   & 17    & \bfseries1.4 & 0.4   & 2.6   & \bfseries129 & 36    & 236 \\
            Ghana & 0.3   & 0.0   & 0.6   & 29    & -5    & 59    & 0.8   & -0.1  & 1.6   & 71    & -6    & 151 \\
            Indonesia & 0.1   & -0.5  & 0.5   & 13    & -58   & 60    & \bfseries0.8 & 0.2   & 1.5   & \bfseries87 & 23    & 166 \\
            Pakistan & 0.1   & -0.2  & 0.4   & 12    & -24   & 44    & \bfseries0.8 & 0.1   & 1.5   & \bfseries88 & 6     & 174 \\
            Philippines & 0.0   & -0.5  & 0.3   & 8     & -76   & 55    & \bfseries0.6 & 0.1   & 1.2   & \bfseries92 & 11    & 198 \\
            Jordan & \bfseries0.5 & 0.2   & 0.7   & \bfseries86 & 30    & 134   & 0.1   & -0.3  & 0.5   & 14    & -59   & 94 \\
            Cameroon &\bfseries0.5 & 0.2   & 0.8   & \bfseries111 & 41    & 181   & -0.1  & -0.8  & 0.8   & -11   & -179  & 172 \\
            Zimbabwe & 0.3   & -0.3  & 0.8   & 92    & -93   & 260   & 0.0   & -0.8  & 0.9   & 8     & -257  & 298 \\
            Colombia & \bfseries0.3 & 0.0   & 0.5   & \bfseries131 & 2     & 267   & -0.1  & -0.6  & 0.5   & -31   & -282  & 251 \\
			\hline
		\end{tabular}
	\end{table}

	\begin{landscape}
		\begin{table}
			\centering
			Table 4. The coefficient by coefficient decomposition results per 1000 births per year for 42 countries. Countries are ordered as in Tables 2 and 3. Bolded results are significant defined as zero not in the 95\% confidence interval.
			\label{tab:betabybeta}
			\vspace{0.1cm}
			\singlespacing
			\footnotesize
			\sisetup{
				table-align-uncertainty=true,
				separate-uncertainty=true,
			}
			%% local redefinitions
			\renewrobustcmd{\bfseries}{\fontseries{b}\selectfont}
			\renewrobustcmd{\boldmath}{}
			\begin{tabular}{
					c
					S[table-format = 1.1, detect-weight,mode=text]
					>{{[}} % Add square bracket before column
					S[table-format = -1.1,table-space-text-pre={[}]
					@{,\,} % Add comma and thin-space between the columns
					S[table-format = -1.2,table-space-text-post={]}]
					<{{]}} % Add square bracket after column
					S[table-format = 1.1, detect-weight,mode=text]
					>{{[}} % Add square bracket before column
					S[table-format = -2,table-space-text-pre={[}]
					@{,\,} % Add comma and thin-space between the columns
					S[table-format = -1.2,table-space-text-post={]}]
					<{{]}} % Add square bracket after column
					S[table-format = 1.1, detect-weight,mode=text]
					>{{[}} % Add square bracket before column
					S[table-format = -1.1,table-space-text-pre={[}]
					@{,\,} % Add comma and thin-space between the columns
					S[table-format = -1.2,table-space-text-post={]}]
					<{{]}} % Add square bracket after column
					S[table-format = 1.1, detect-weight,mode=text]
					>{{[}} % Add square bracket before column
					S[table-format = -1.1,table-space-text-pre={[}]
					@{,\,} % Add comma and thin-space between the columns
					S[table-format = -1.2,table-space-text-post={]}]
					<{{]}} % Add square bracket after column
				}
				\hline
				Country  &{Overall} & \multicolumn{2}{p{2cm}}{95\% CI}  & {Intercept} & \multicolumn{2}{p{2cm}}{95\% CI} &{Wealth score}& \multicolumn{2}{p{2cm}}{95\% CI} &{Maternal edu} & \multicolumn{2}{p{2cm}}{95\% CI} \cr
				\hline
				      Cambodia & \textbf{4.4} & 3.5   & 5.3   & 3.9   & -1.2  & 6.3   & -0.9  & -3.5  & 1.8   & 0.6   & -0.3  & 2.0 \\
				      Niger & \textbf{4.9} & 3.8   & 6.0   & 4.4   & -1.4  & 7.9   & 0.1   & -3.0  & 4.2   & 0.0   & -0.3  & 0.2 \\
				      Mali  & \textbf{4.6} & 3.6   & 5.6   & \textbf{6.2} & 2.6   & 8.3   & 0.2   & -1.5  & 2.7   & 0.0   & -0.1  & 0.1 \\
				      Madagascar & \textbf{3.8} & 2.6   & 5.1   & 3.2   & -2.5  & 6.8   & 0.0   & -3.1  & 4.2   & -0.3  & -1.3  & 0.7 \\
				      Malawi & \textbf{4.4} & 3.4   & 5.3   & 2.4   & -2.2  & 5.3   & 0.4   & -1.8  & 3.4   & 0.4   & -0.6  & 1.5 \\
				      Zambia & \textbf{4.0} & 3.1   & 4.9   & -0.8  & -9.5  & 4.6   & 2.3   & -1.5  & 7.8   & -0.6  & -2.3  & 1.0 \\
				      Benin & \textbf{3.6} & 2.6   & 4.7   & \textbf{5.2} & 2.4   & 7.0   & -0.4  & -1.9  & 1.1   & -0.1  & -0.3  & 0.1 \\
				      Mozambique & \textbf{3.4} & 2.3   & 4.5   & 0.1   & -7.4  & 5.6   & 1.3   & -3.0  & 6.9   & 0.0   & -1.0  & 1.2 \\
				      Comoros & \textbf{2.9} & 1.5   & 4.6   & \textbf{3.8} & 0.8   & 5.7   & 0.1   & -1.0  & 1.8   & -0.1  & -0.6  & 0.4 \\
				      Tanzania & \textbf{3.4} & 2.4   & 4.4   & \textbf{4.8} & 1.9   & 6.4   & -0.3  & -1.6  & 1.3   & \bfseries-0.9 & -2.0  & -0.2 \\
				      Cote dIvoire & \textbf{3.4} & 1.1   & 6.1   & 4.4   & -2.9  & 9.2   & 1.7   & -1.3  & 6.7   & -0.1  & -0.6  & 0.4 \\
				      Chad  & \textbf{2.3} & 1.1   & 3.5   & 2.3   & -1.9  & 5.3   & 0.3   & -2.1  & 3.6   & \bfseries-0.4 & -0.7  & -0.1 \\
				      Armenia & 1.5   & -0.1  & 3.0   & -0.9  & -13.0 & 3.4   & 0.6   & -2.4  & 5.1   & 1.0   & -2.7  & 7.3 \\
				      Rwanda & \textbf{2.3} & 1.6   & 3.0   & 0.3   & -4.6  & 3.2   & 1.4   & -0.5  & 4.8   & 0.0   & -0.8  & 0.7 \\
				      Egypt & \textbf{1.6} & 1.1   & 2.2   & 1.2   & -2.2  & 2.7   & 0.5   & -0.4  & 2.6   & 0.1   & -0.3  & 0.6 \\
				      Turkey & \textbf{2.2} & 1.0   & 3.6   & 2.3   & -3.4  & 5.2   & 0.8   & -1.1  & 4.1   & 0.0   & -0.8  & 0.9 \\
				      Guinea & \textbf{2.1} & 0.8   & 3.4   & 0.7   & -7.8  & 5.8   & 0.4   & -3.8  & 6.3   & 0.1   & -0.3  & 0.6 \\
				      Togo  & \textbf{1.9} & 1.0   & 2.9   & 1.8   & -3.4  & 4.8   & 0.3   & -2.2  & 3.4   & 0.4   & -0.1  & 1.2 \\
				      Bolivia & \textbf{1.8} & 0.5   & 3.2   & -3.0  & -14.8 & 3.9   & -1.8  & -7.4  & 3.9   & \bfseries3.6 & 0.0   & 8.7 \\
				      Bangladesh & \textbf{1.5} & 0.7   & 2.4   & 2.3   & -1.0  & 4.1   & 0.7   & -0.8  & 3.1   & -0.3  & -1.1  & 0.2 \\
				      Kyrgyzstan & \textbf{1.9} & 0.5   & 3.7   & 2.0   & -4.3  & 5.0   & -0.2  & -2.6  & 2.7   & -0.3  & -3.5  & 3.9 \\
				      Senegal & \textbf{2.2} & 1.4   & 3.0   & 0.3   & -5.4  & 3.6   & 1.4   & -0.9  & 5.1   & 0.0   & -0.3  & 0.3 \\
				      Peru  & \textbf{1.2} & 0.8   & 1.6   & 0.0   & -3.1  & 1.7   & \textbf{1.1} & 0.1   & 3.4   & 0.0   & -0.5  & 0.6 \\
				      Burkina Faso & \textbf{2.1} & 1.2   & 3.2   & 3.8   & -0.6  & 6.3   & -0.2  & -2.2  & 2.2   & 0.1   & -0.1  & 0.2 \\
				      Uganda & \textbf{1.7} & 0.8   & 2.5   & \textbf{3.6} & 0.4   & 5.4   & 0.2   & -1.2  & 2.2   & \bfseries-0.8 & -1.7  & -0.2 \\
				      Gabon & \textbf{2.0} & 0.8   & 3.5   & 0.7   & -7.8  & 5.0   & 1.5   & -1.5  & 6.9   & 0.3   & -1.6  & 3.3 \\
				      Morocco & \textbf{1.4} & 0.0   & 2.9   & 0.7   & -7.6  & 4.9   & 0.4   & -3.0  & 4.9   & 0.2   & -0.3  & 1.0 \\
				      Guatemala & \textbf{1.0} & 0.3   & 1.7   & 1.8   & -0.2  & 3.0   & -0.2  & -1.1  & 0.9   & -0.1  & -0.6  & 0.3 \\
				      India & \textbf{1.3} & 1.0   & 1.6   & 1.2   & -1.4  & 3.3   & 0.5   & -0.8  & 2.2   & 0.0   & -0.3  & 0.2 \\
				      Nigeria & \textbf{0.9} & 0.3   & 1.5   & 0.8   & -2.7  & 3.1   & 0.3   & -1.4  & 2.6   & -0.1  & -0.5  & 0.2 \\
				      Namibia & \textbf{0.9} & 0.0   & 1.9   & -0.3  & -5.8  & 2.7   & 0.7   & -1.6  & 4.0   & -0.7  & -2.4  & 0.9 \\
				      Kenya & \textbf{1.2} & 0.6   & 1.9   & \textbf{2.3} & 0.6   & 3.2   & -0.1  & -0.8  & 0.7   & 0.1   & -0.2  & 0.6 \\
				      Haiti & 1.0   & -0.2  & 2.3   & 0.8   & -5.4  & 4.5   & 0.8   & -1.6  & 4.7   & 0.0   & -1.0  & 1.1 \\
				      Dominican Republic & \textbf{1.4} & 0.4   & 2.6   & -4.7  & -14.5 & 1.7   & 1.5   & -2.7  & 7.1   & 2.0   & -1.3  & 6.9 \\
				      Ghana & 0.8   & -0.1  & 1.6   & 0.1   & -4.9  & 2.8   & -0.8  & -3.5  & 1.9   & 0.5   & -0.4  & 1.6 \\
				      Indonesia & \textbf{0.8} & 0.2   & 1.5   & -2.3  & -8.8  & 1.5   & 0.9   & -1.7  & 4.7   & 0.6   & -0.9  & 2.6 \\
				      Pakistan & \textbf{0.8} & 0.1   & 1.5   & 1.1   & -3.5  & 3.5   & 1.2   & -0.5  & 4.3   & 0.0   & -0.3  & 0.3 \\
				      Philippines & \textbf{0.6} & 0.1   & 1.2   & 0.3   & -3.2  & 1.9   & 0.2   & -1.1  & 2.0   & -0.5  & -1.8  & 0.8 \\
				      Jordan & 0.1   & -0.3  & 0.5   & -0.2  & -3.0  & 1.0   & -0.6  & -2.4  & 0.6   & -0.9  & -2.8  & 0.4 \\
				      Cameroon & -0.1  & -0.8  & 0.8   & 0.4   & -4.3  & 2.9   & 0.8   & -1.0  & 3.8   & -0.9  & -1.9  & -0.2 \\
				      Zimbabwe & 0.0   & -0.8  & 0.9   & -1.6  & -7.8  & 1.9   & 0.5   & -2.1  & 4.6   & 0.4   & -1.8  & 3.2 \\
				      Colombia & -0.1  & -0.6  & 0.5   & -1.4  & -7.0  & 1.1   & -0.5  & -3.3  & 1.8   & -0.4  & -2.8  & 2.0 \\
				\hline
			\end{tabular}
		\end{table}
	\end{landscape}
	
	\setcounter{table}{1}
	
	\begin{landscape}
		\begin{table}
		\centering
			Table 4 (Continued)
			\label{tab:betabybeta2}
			\vspace{0.1cm}
			\singlespacing
			\footnotesize
			\sisetup{
				table-align-uncertainty=true,
				separate-uncertainty=true,
			}
			%% local redefinitions
			\renewrobustcmd{\bfseries}{\fontseries{b}\selectfont}
			\renewrobustcmd{\boldmath}{}
		\hspace*{-1.3cm}\begin{tabular}{
					c
					S[table-format = 1.1, detect-weight,mode=text]
					>{{[}} % Add square bracket before column
					S[table-format = -1.1,table-space-text-pre={[}]
					@{,\,} % Add comma and thin-space between the columns
					S[table-format = -1.2,table-space-text-post={]}]
					<{{]}} % Add square bracket after column
					S[table-format = 1.1, detect-weight,mode=text]
					>{{[}} % Add square bracket before column
					S[table-format = -1.1,table-space-text-pre={[}]
					@{,\,} % Add comma and thin-space between the columns
					S[table-format = -1.2,table-space-text-post={]}]
					<{{]}} % Add square bracket after column
					S[table-format = 1.1, detect-weight,mode=text]
					>{{[}} % Add square bracket before column
					S[table-format = -1.1,table-space-text-pre={[}]
					@{,\,} % Add comma and thin-space between the columns
					S[table-format = -1.2,table-space-text-post={]}]
					<{{]}} % Add square bracket after column
					S[table-format = 1.1, detect-weight,mode=text]
					>{{[}} % Add square bracket before column
					S[table-format = -1.1,table-space-text-pre={[}]
					@{,\,} % Add comma and thin-space between the columns
					S[table-format = -1.2,table-space-text-post={]}]
					<{{]}} % Add square bracket after column
					S[table-format = 1.1, detect-weight,mode=text]
					>{{[}} % Add square bracket before column
					S[table-format = -1.1,table-space-text-pre={[}]
					@{,\,} % Add comma and thin-space between the columns
					S[table-format = -1.2,table-space-text-post={]}]
					<{{]}} % Add square bracket after column			
				}
				\hline
				Country  &{Maternal age}& \multicolumn{2}{p{2cm}}{95\% CI}  & {Birth order} & \multicolumn{2}{p{2cm}}{95\% CI} &{Birth Interval}& \multicolumn{2}{p{2cm}}{95\% CI}&{Sex}& \multicolumn{2}{p{2cm}}{95\% CI}&{Residence}& \multicolumn{2}{p{2cm}}{95\% CI}\cr
				\hline
				         Cambodia & 0.5   & -1.6  & 4.2   & -0.2  & -1.8  & 0.9   & 0.0   & -1.1  & 1.8   & 0.1   & -0.4  & 0.6   & \bfseries0.4 & 0.1   & 0.7 \\
				     Niger & 0.0   & -2.9  & 3.5   & -2.1  & -7.4  & 1.6   & 2.6   & -1.1  & 8.2   & -0.2  & -0.7  & 0.3   & 0.1   & -0.3  & 0.5 \\
				     Mali  & 0.0   & -1.3  & 1.9   & -0.1  & -2.0  & 1.0   & -0.9  & -2.0  & 1.0   & \bfseries-0.6 & -1.0  & -0.2  & -0.2  & -0.4  & 0.1 \\
				     Madagascar & 0.0   & -2.9  & 3.4   & -1.0  & -5.5  & 2.3   & 1.8   & -1.2  & 6.4   & 0.0   & -0.6  & 0.6   & 0.0   & -0.3  & 0.4 \\
				     Malawi & 1.0   & -0.7  & 3.5   & -0.9  & -3.0  & 0.7   & 1.2   & -0.3  & 3.3   & -0.1  & -0.4  & 0.2   & 0.0   & -0.1  & 0.2 \\
				     Zambia & 1.3   & -2.4  & 6.1   & 0.4   & -2.6  & 3.0   & 1.1   & -1.0  & 4.5   & 0.2   & -0.2  & 0.7   & 0.1   & -0.5  & 0.8 \\
				     Benin & 0.3   & -1.1  & 2.6   & -0.8  & -2.6  & 0.4   & -0.3  & -1.5  & 1.7   & -0.2  & -0.5  & 0.2   & -0.2  & -0.6  & 0.2 \\
				     Mozambique & 2.3   & -0.9  & 6.6   & -1.6  & -5.6  & 1.2   & 1.4   & -1.4  & 5.7   & 0.0   & -0.5  & 0.6   & -0.3  & -0.8  & 0.4 \\
				     Comoros & -0.7  & -2.2  & 0.9   & -0.7  & -3.1  & 0.8   & 0.4   & -1.0  & 3.0   & 0.4   & -0.2  & 1.1   & -0.2  & -0.6  & 0.4 \\
				     Tanzania & 0.4   & -1.3  & 3.3   & -0.7  & -2.8  & 0.7   & 0.3   & -1.0  & 2.6   & -0.3  & -0.7  & 0.2   & -0.1  & -0.4  & 0.4 \\
				     Cote dIvoire & -1.9  & -4.8  & 1.6   & -2.0  & -7.2  & 1.7   & 1.7   & -1.9  & 6.9   & -0.8  & -2.0  & 0.4   & 0.5   & -0.6  & 2.1 \\
				     Chad  & 0.4   & -1.5  & 2.9   & -0.8  & -4.4  & 1.5   & 0.6   & -1.7  & 4.2   & -0.3  & -0.7  & 0.1   & 0.2   & -0.2  & 0.7 \\
				     Armenia & 0.0   & -2.4  & 3.9   & 0.2   & -1.4  & 1.8   & 0.2   & -0.9  & 1.9   & 0.3   & -0.6  & 1.4   & 0.2   & -1.0  & 2.1 \\
				     Rwanda & 0.3   & -1.1  & 3.2   & 0.2   & -0.9  & 1.0   & -0.2  & -0.8  & 0.9   & 0.2   & -0.1  & 0.5   & 0.1   & -0.1  & 0.4 \\
				     Egypt & -0.3  & -1.2  & 1.3   & -0.7  & -2.2  & 0.3   & \bfseries0.9 & 0.0   & 2.4   & -0.1  & -0.3  & 0.1   & 0.0   & -0.2  & 0.3 \\
				     Turkey & -0.2  & -2.0  & 2.7   & 0.1   & -1.7  & 1.3   & -1.0  & -2.5  & 0.9   & 0.1   & -0.6  & 1.0   & 0.1   & -1.0  & 1.6 \\
				     Guinea & 0.1   & -3.4  & 4.7   & -1.5  & -7.0  & 2.6   & 1.9   & -2.0  & 7.7   & -0.1  & -0.9  & 0.8   & 0.3   & -0.5  & 1.4 \\
				     Togo  & -0.8  & -3.0  & 2.7   & -1.0  & -4.4  & 1.6   & 0.8   & -1.6  & 4.3   & 0.2   & -0.4  & 0.7   & 0.2   & -0.3  & 0.9 \\
				     Bolivia &\bfseries4.5 & 0.0   & 11.7  & -0.4  & -3.6  & 1.6   & 0.0   & -1.9  & 3.2   & -0.4  & -1.0  & 0.2   & -0.6  & -1.3  & 0.3 \\
				     Bangladesh & -0.6  & -1.8  & 1.0   & -0.6  & -2.8  & 0.8   & 0.2   & -1.1  & 2.4   & -0.2  & -0.6  & 0.3   & 0.1   & -0.3  & 0.6 \\
				     Kyrgyzstan & -0.4  & -2.5  & 3.3   & 0.0   & -2.1  & 1.8   & 0.2   & -1.2  & 2.4   & 0.6   & -0.1  & 1.5   & -0.1  & -0.4  & 0.4 \\
				     Senegal & 0.2   & -1.9  & 2.9   & -0.4  & -3.1  & 1.4   & 1.1   & -0.5  & 3.9   & \bfseries-0.4 & -0.7  & -0.1  & -0.1  & -0.3  & 0.2 \\
				     Peru  & 0.2   & -0.6  & 1.2   & -0.5  & -1.7  & 0.3   & 0.5   & -0.3  & 1.8   & 0.0   & -0.1  & 0.2   & -0.1  & -0.3  & 0.1 \\
				     Burkina Faso & 0.1   & -1.8  & 3.2   & -0.6  & -3.1  & 1.0   & -0.9  & -2.6  & 1.6   & 0.0   & -0.5  & 0.5   & -0.2  & -0.5  & 0.1 \\
				     Uganda & -0.5  & -2.2  & 1.8   & -0.5  & -3.4  & 1.5   & 0.1   & -1.8  & 3.4   & \bfseries-0.6 & -1.0  & -0.1  & 0.2   & -0.2  & 0.6 \\
				     Gabon & -0.9  & -3.7  & 2.3   & 1.0   & -1.4  & 3.3   & -1.5  & -3.2  & 0.9   & \bfseries1.0 & 0.1   & 2.1   & -0.1  & -1.3  & 1.4 \\
				     Morocco & 0.0   & -3.2  & 5.7   & -0.2  & -3.3  & 2.1   & 0.2   & -1.8  & 3.5   & -0.1  & -0.9  & 0.8   & 0.1   & -0.7  & 1.3 \\
				     Guatemala & -0.5  & -1.6  & 0.9   & -0.1  & -1.8  & 0.9   & 0.4   & -0.6  & 2.0   & \bfseries-0.3 & -0.6  & 0.0   & 0.0   & -0.3  & 0.4 \\
				     India & -0.3  & -1.6  & 1.6   & \bfseries1.1 & 0.0   & 1.9   & \bfseries-1.2 & -1.9  & -0.1  & -0.2  & -0.4  & 0.1   & 0.0   & -0.2  & 0.2 \\
				     Nigeria & -0.4  & -2.2  & 1.7   & -1.8  & -4.9  & 0.5   & 1.9   & -0.4  & 5.0   & \bfseries0.3 & 0.0   & 0.7   & -0.1  & -0.4  & 0.2 \\
				     Namibia & 0.2   & -2.1  & 3.4   & -1.2  & -4.0  & 0.9   & 1.9   & -0.1  & 4.8   & 0.0   & -0.5  & 0.5   & 0.2   & -0.5  & 1.0 \\
				     Kenya & -0.3  & -1.0  & 0.8   & -0.8  & -2.5  & 0.2   & 0.3   & -0.8  & 2.0   & -0.1  & -0.4  & 0.2   & -0.1  & -0.4  & 0.3 \\
				     Haiti & 0.0   & -2.3  & 3.6   & -0.6  & -3.1  & 1.1   & 0.4   & -1.2  & 3.0   & 0.0   & -0.6  & 0.6   & -0.4  & -1.1  & 0.5 \\
				     Dominican Republic & 1.4   & -1.5  & 5.5   & -0.4  & -2.6  & 1.5   & 0.9   & -0.7  & 3.4   & 0.2   & -0.5  & 1.0   & 0.6   & -0.3  & 1.9 \\
				     Ghana & 0.4   & -2.0  & 4.2   & -0.1  & -2.4  & 1.9   & 0.4   & -1.3  & 2.9   & 0.4   & -0.1  & 1.1   & -0.2  & -0.8  & 0.5 \\
				     Indonesia & 1.1   & -1.3  & 5.2   & -0.4  & -2.6  & 1.0   & 0.8   & -0.6  & 3.0   & 0.2   & -0.1  & 0.5   & 0.0   & -0.3  & 0.3 \\
				     Pakistan & -0.7  & -2.2  & 1.6   & -0.4  & -2.8  & 1.1   & -0.3  & -1.8  & 2.1   & 0.2   & -0.2  & 0.7   & -0.2  & -0.6  & 0.2 \\
				     Philippines & 0.1   & -1.4  & 2.8   & 0.7   & -0.2  & 1.6   & -0.3  & -0.9  & 0.6   & 0.1   & -0.2  & 0.4   & -0.1  & -0.3  & 0.2 \\
				     Jordan & 1.0   & -1.0  & 4.8   & -0.4  & -2.4  & 0.9   &\bfseries1.3 & 0.1   & 3.3   & -0.1  & -0.3  & 0.1   & 0.0   & -0.3  & 0.5 \\
				     Cameroon & -0.3  & -2.4  & 2.6   & -1.4  & -4.6  & 0.8   & 0.6   & -1.6  & 3.8   & 0.3   & -0.4  & 1.1   & 0.6   & -0.3  & 1.7 \\
				     Zimbabwe & 1.0   & -1.1  & 4.4   & -0.6  & -2.7  & 0.9   & 0.3   & -1.0  & 2.6   & 0.0   & -0.4  & 0.6   & -0.1  & -0.7  & 1.0 \\
				     Colombia & 1.0   & -1.4  & 4.9   & -1.4  & -4.6  & 0.4   & 1.5   & -0.4  & 5.0   & 0.3   & -0.4  & 1.1   & 0.9   & -0.3  & 2.8 \\
				\hline
			\end{tabular}\hspace*{-1cm}
		\end{table}
	\end{landscape}

\section*{Web Appendix for \\``Explaining the Decline of Child Mortality in 44 Developing Countries: \\A Bayesian Extension of Oaxaca Decomposition Methods"}

\clearpage

	\subsection*{A.1 Summary of Categorical Covariates}
	\begin{table}[H]
		\centering
		Table 5. Proportion of Male and Urban births in each survey for 42 countries
		\vspace{0.1cm}
		\singlespacing
		\footnotesize
		\begin{tabular}{ccccc}
			\hline
			&\multicolumn{2}{c}{Survey1}&\multicolumn{2}{c}{Survey2}\\
			Country &Male& Urban&Male& Urban \\
			\hline
			Armenia & 0.54  & 0.54  & 0.43  & 0.67 \\
			Bangladesh & 0.51  & 0.51  & 0.24  & 0.32 \\
			Benin & 0.51  & 0.51  & 0.27  & 0.35 \\
			Bolivia & 0.51  & 0.52  & 0.52  & 0.51 \\
			Burkina Faso & 0.51  & 0.51  & 0.31  & 0.21 \\
			Cambodia & 0.51  & 0.50  & 0.14  & 0.27 \\
			Cameroon & 0.51  & 0.50  & 0.53  & 0.40 \\
			Chad  & 0.51  & 0.51  & 0.37  & 0.20 \\
			Colombia & 0.49  & 0.51  & 0.82  & 0.69 \\
			Comoros & 0.51  & 0.50  & 0.24  & 0.33 \\
			Cote dIvoire & 0.50  & 0.50  & 0.54  & 0.32 \\
			Dominican Republic & 0.51  & 0.51  & 0.52  & 0.56 \\
			Egypt & 0.52  & 0.52  & 0.34  & 0.41 \\
			Gabon & 0.52  & 0.50  & 0.60  & 0.61 \\
			Ghana & 0.51  & 0.52  & 0.27  & 0.40 \\
			Guatemala & 0.51  & 0.52  & 0.23  & 0.35 \\
			Guinea & 0.52  & 0.52  & 0.27  & 0.29 \\
			Haiti & 0.50  & 0.51  & 0.38  & 0.34 \\
			India & 0.52  & 0.52  & 0.28  & 0.38 \\
			Indonesia & 0.51  & 0.52  & 0.26  & 0.46 \\
			Jordan & 0.52  & 0.51  & 0.66  & 0.70 \\
			Kenya & 0.50  & 0.51  & 0.11  & 0.33 \\
			Kyrgyzstan & 0.50  & 0.51  & 0.28  & 0.26 \\
			Madagascar & 0.51  & 0.52  & 0.23  & 0.17 \\
			Malawi & 0.51  & 0.50  & 0.26  & 0.17 \\
			Mali  & 0.51  & 0.52  & 0.30  & 0.23 \\
			Morocco & 0.51  & 0.51  & 0.35  & 0.43 \\
			Mozambique & 0.50  & 0.51  & 0.26  & 0.33 \\
			Namibia & 0.48  & 0.50  & 0.30  & 0.47 \\
			Niger & 0.50  & 0.51  & 0.25  & 0.22 \\
			Nigeria & 0.51  & 0.51  & 0.35  & 0.33 \\
			Pakistan & 0.51  & 0.51  & 0.52  & 0.42 \\
			Peru  & 0.52  & 0.51  & 0.58  & 0.56 \\
			Philippines & 0.51  & 0.52  & 0.45  & 0.41 \\
			Rwanda & 0.50  & 0.51  & 0.14  & 0.22 \\
			Senegal & 0.52  & 0.50  & 0.27  & 0.28 \\
			Tanzania & 0.51  & 0.51  & 0.24  & 0.22 \\
			Togo  & 0.51  & 0.50  & 0.22  & 0.27 \\
			Turkey & 0.51  & 0.51  & 0.61  & 0.67 \\
			Uganda & 0.49  & 0.51  & 0.28  & 0.21 \\
			Zambia & 0.49  & 0.51  & 0.34  & 0.37 \\
			Zimbabwe & 0.50  & 0.50  & 0.22  & 0.40 \\
			\hline
		\end{tabular}%
		\label{tab:addlabel}%
	\end{table}%

	\subsection*{A.2 The Distribution of Five Continous Covariates}
	\begin{figure}[H]
	\includegraphics[width=18cm, height = 18cm]{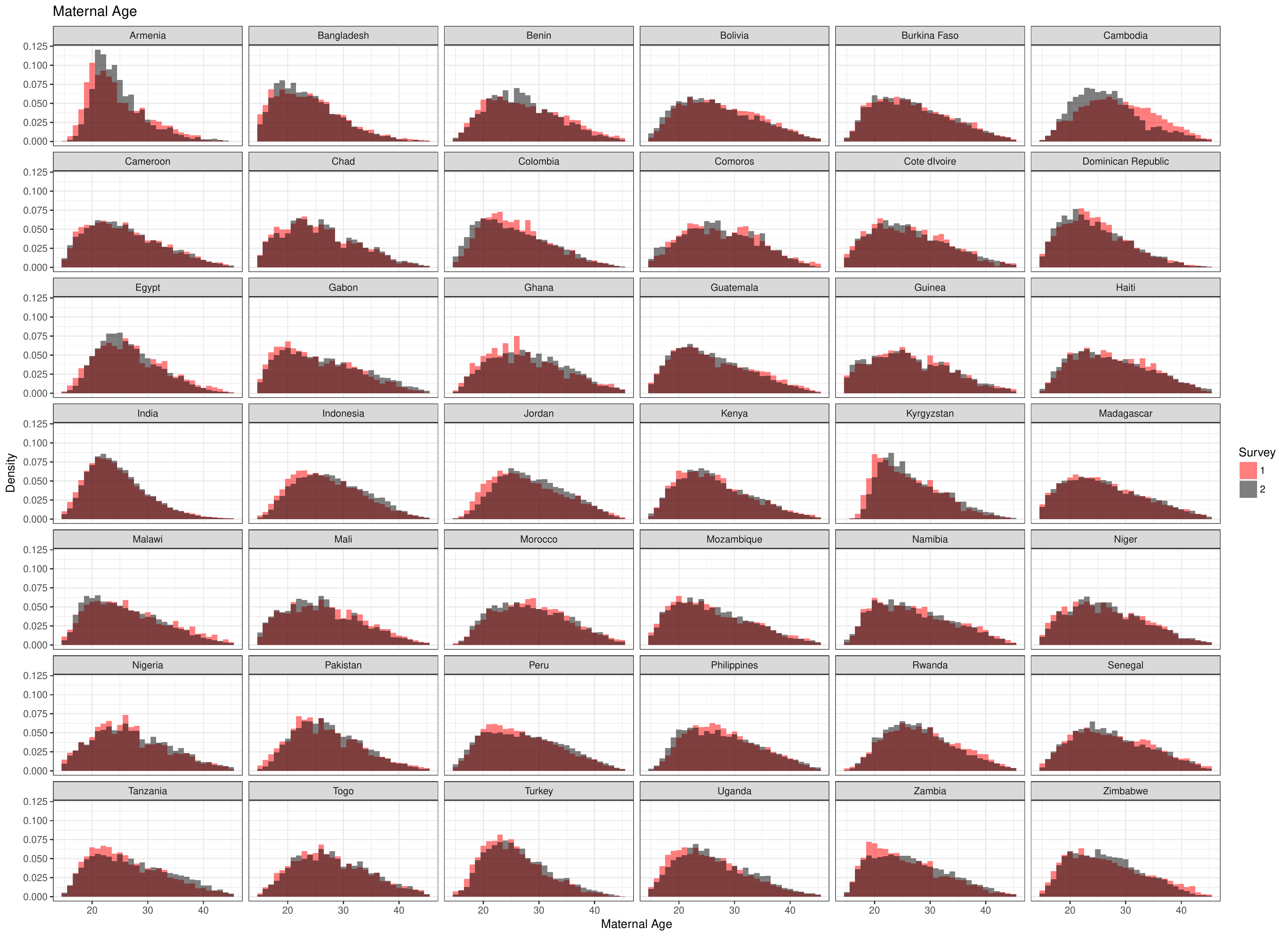}
		\centering
		\caption{Histogram of maternal age in two surveys for 42 countries}
	\end{figure}
	
	\begin{figure}[H]
	
		\includegraphics[width=18cm, height = 18cm]{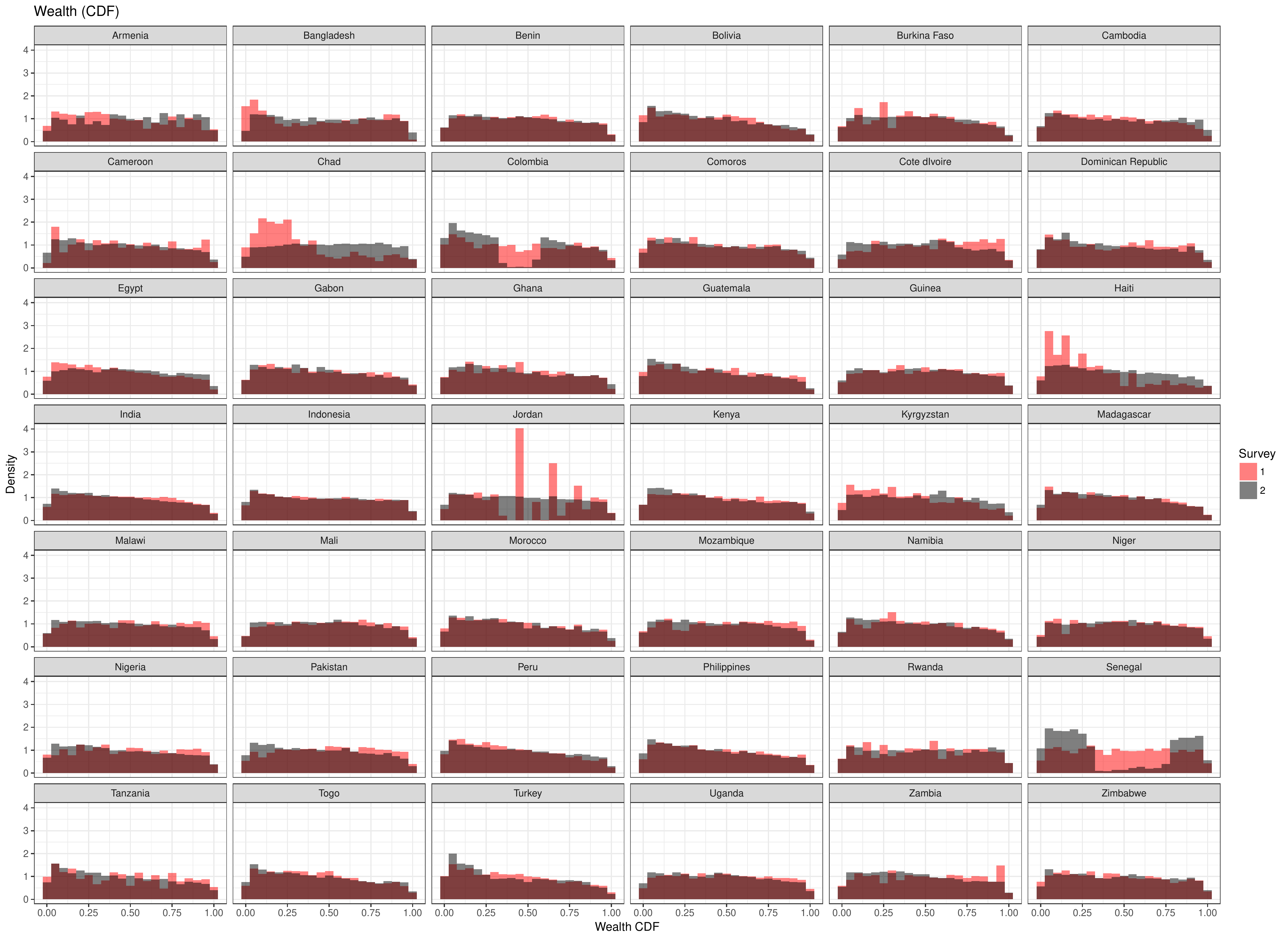}
		\centering
		\caption{Histogram of weath CDF in two surveys for 42 countries}
	\end{figure}
	
	\begin{figure}[H]
		\includegraphics[width=18cm, height = 18cm]{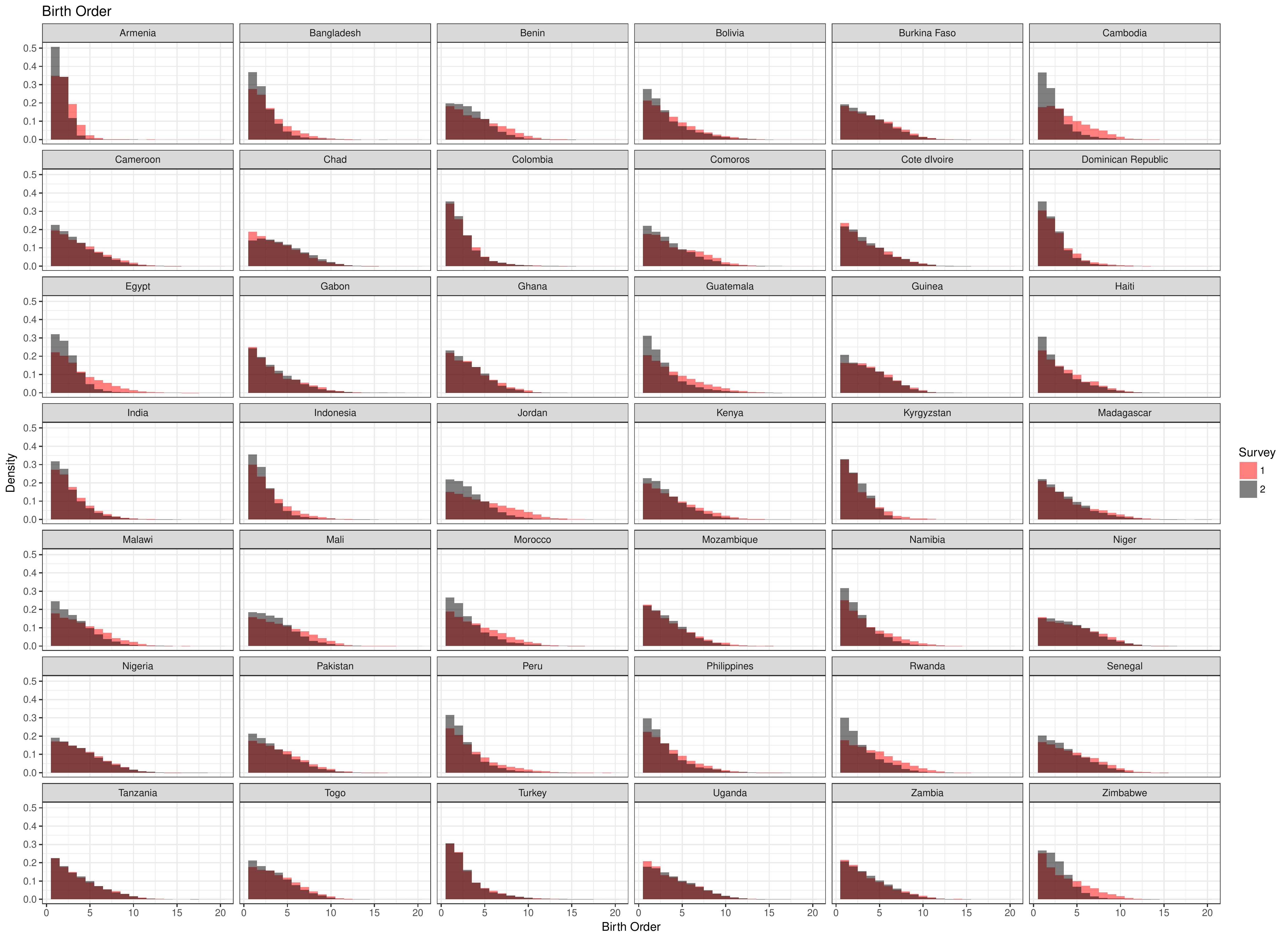}
		\centering
		\caption{Histogram of birth order in two surveys for 42 countries}
	\end{figure}
	
	\begin{figure}[H]
		\includegraphics[width=18cm, height = 18cm]{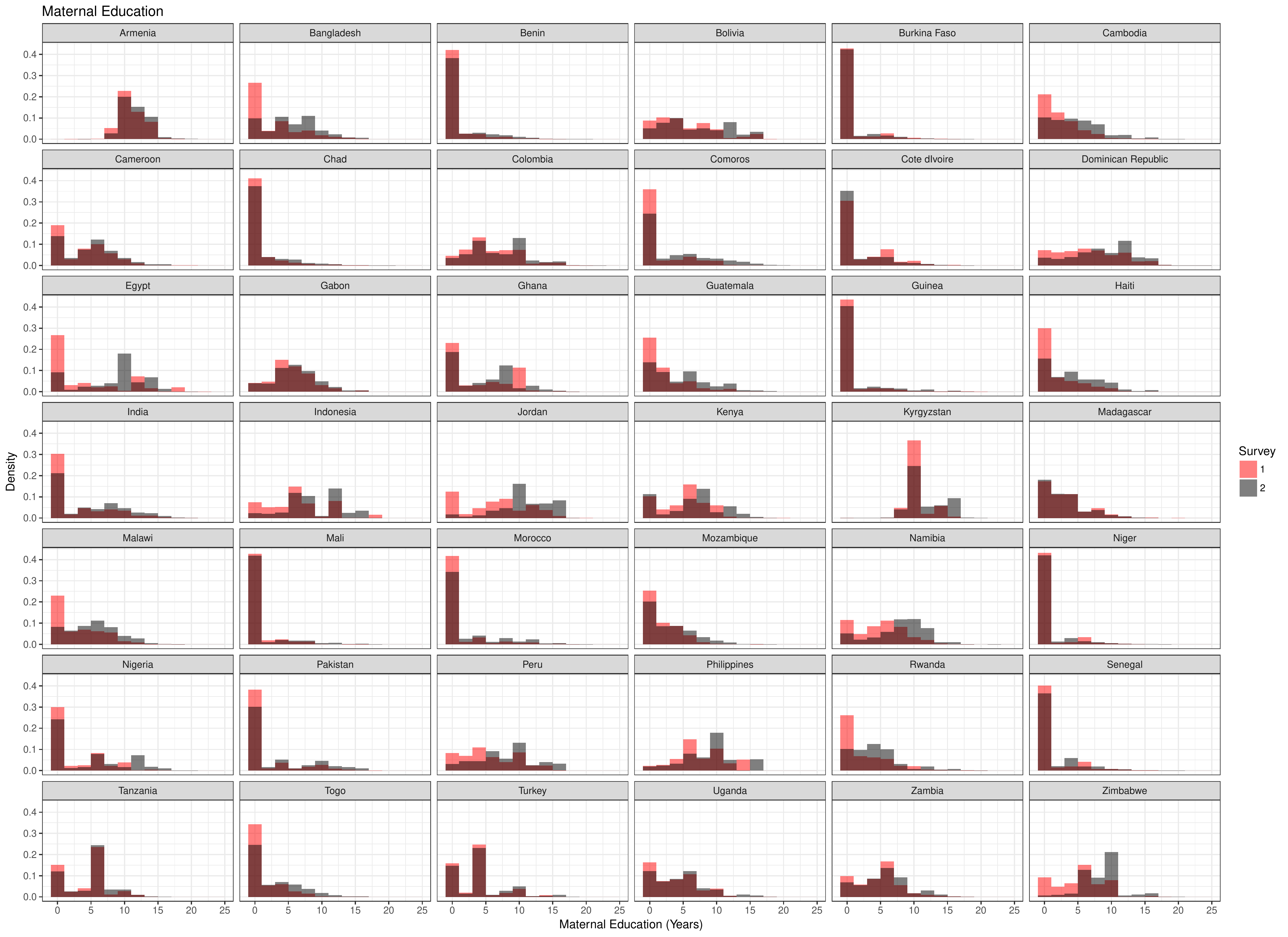}
		\centering
		\caption{Histogram of maternal education in two surveys for 42 countries}
	\end{figure}

\begin{figure}[H]
	\includegraphics[width=18cm, height = 18cm]{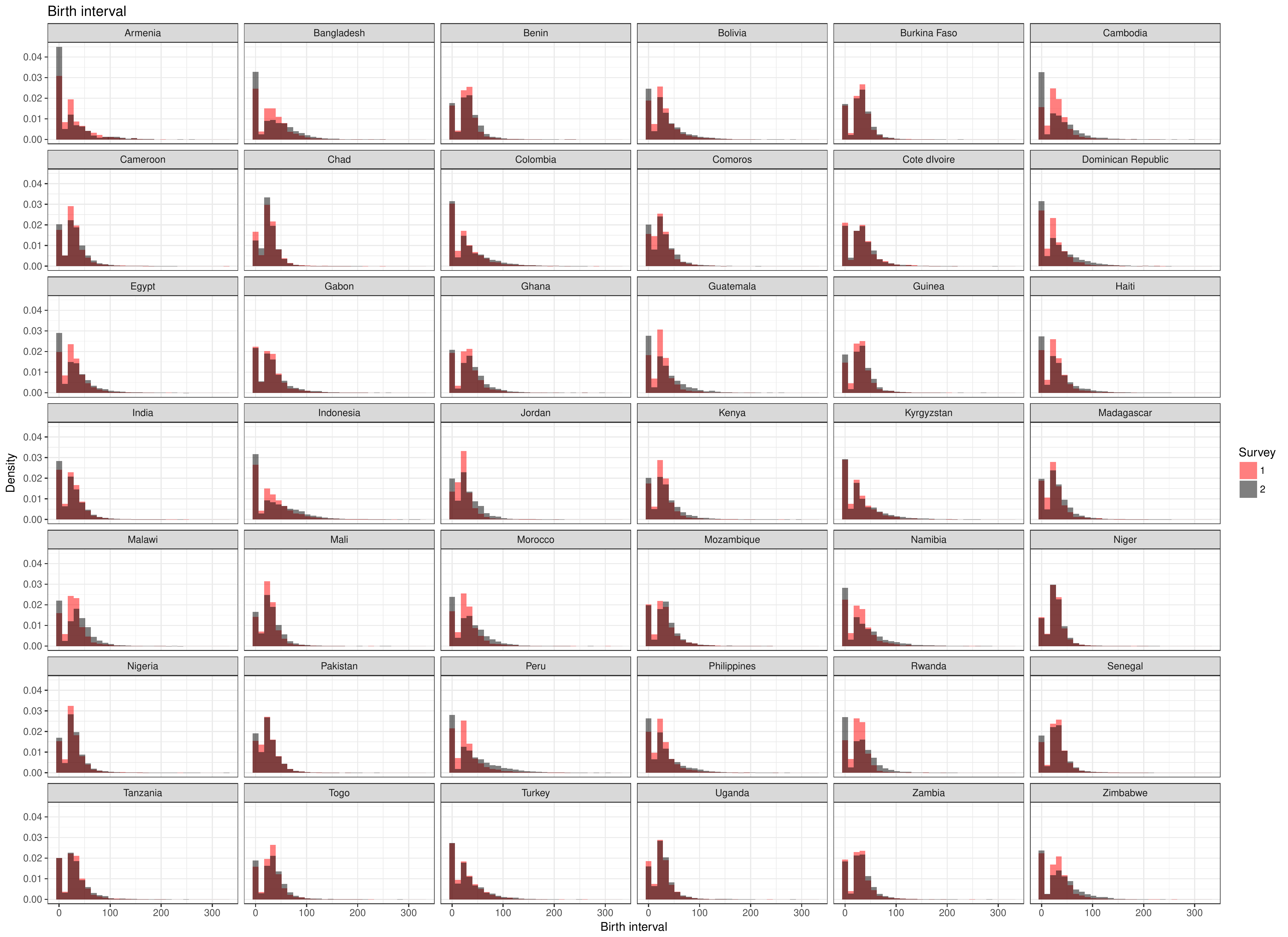}
	\centering
	\caption{Histogram of birth interval in two surveys for 42 countries}
\end{figure}

\subsection*{A.3 Statistical Significance in the Beta-by-Beta Decomposition.}
We investigated the reasons for the lack of statistical significance in the beta by beta decompositions. Table 4 shows that the posterior interval for the intercept effect is always larger than the interval of the overall effect, despite the fact that the intercept term is one term in the sum that makes up the overall effect. In fact, because the overall coefficient effect can be written as a collapsing sum of the individual effects, we can examine how the variance of the overall effect changes as more and more terms are incorporated into the sum. 

\begin{figure}[H]
	\includegraphics[width=18cm, height = 18cm]{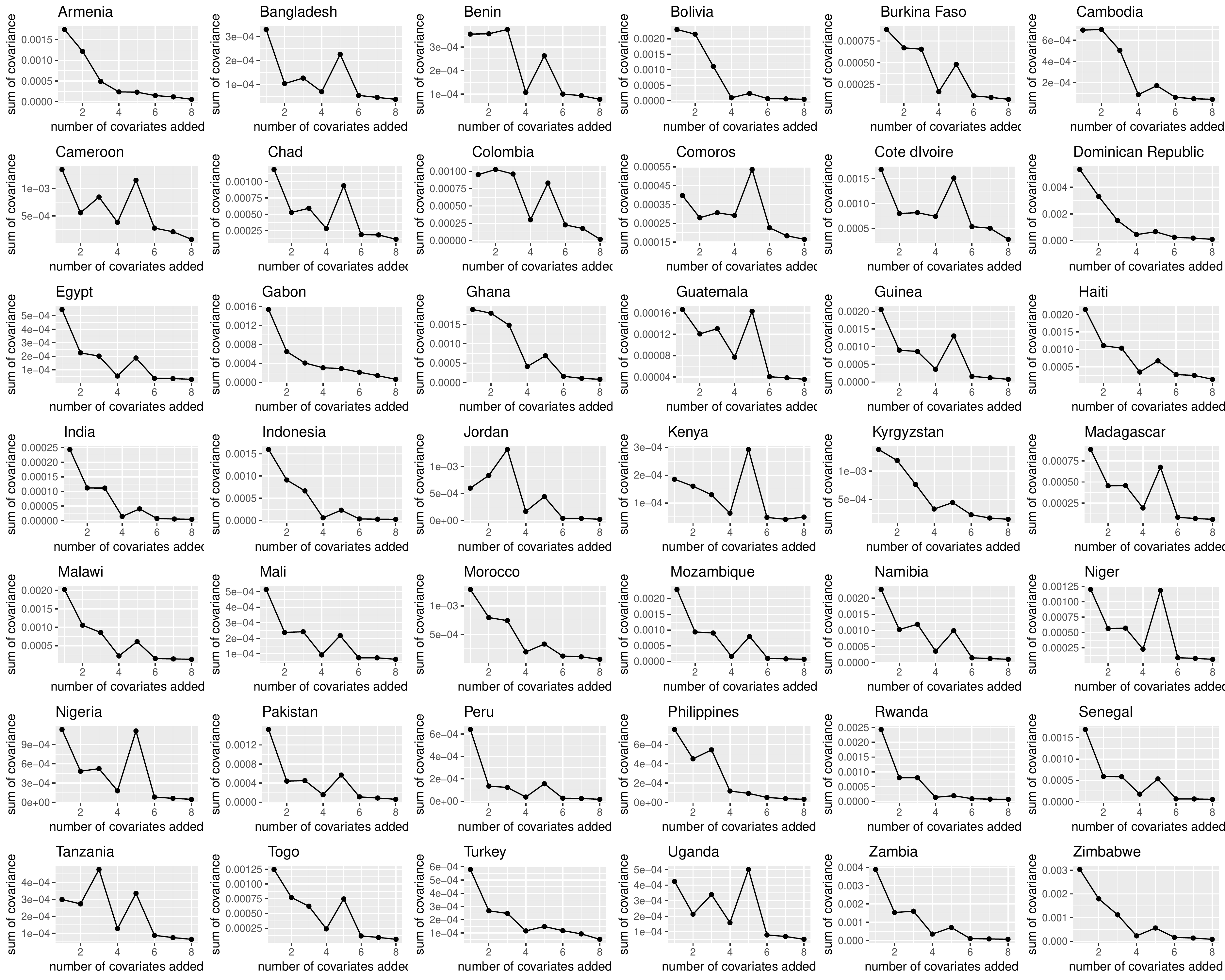}
	\centering
	\caption{The overall coefficient effect \eqref{eq:decomp9} can be written as a collapsing sum of the individual coefficients. The above figure plots the variance as more terms are added to the sum. The variance of the sum remains larger than the variance of the overall coefficient effect until the last coefficient is added into the sum.}
	\label{fig:variance}
\end{figure}

The results of this exercise are plotted in Figure \ref{fig:variance}. First, the countries generally follow one of two trends: Either the variance starts large and gradually decreases until the final term is added to the decomposition, or the variance decreases until the fifth term, birth order, is added to the decomposition which increases the variance. Then the variance decreases again until the last term is added to the sum. Because the models for each country were fit separately, this suggests that the trends we see are not spurious. Second, the variance \emph{never} reaches its minimum until the final term is added to the model. This suggests that the individual components of the sum are highly negatively correlated which we investigated individually for several surveys and found that was indeed the case.

\subsection*{A.4 Integrating Random Effects}

In equations 14 --- 21 we integrated out the cluster level random effects $\gamma_{jk}$ to do the Oaxaca decomposition, allowing us to calculate the marginal expected values of the $y_{ijk}$. For a derivation of this result, we start with the iterated expectation formula. Let $Z$ be a standard normal random variable. Then

\begin{align}
E[y_{ijk}] &= E[E(y_{ijk} | \gamma_{jk})]\\
&= E\Big[\Phi(\bm{x}_{ijk}^{T}\boldsymbol\beta_{k} + \gamma_{jk})\Big]\\
&= E\Big[P(Z < \bm{x}_{ijk}^{T}\boldsymbol\beta_{k} + \gamma_{jk} | \gamma_{jk})\Big] \\
&= P(Z < \bm{x}_{ijk}^{T}\boldsymbol\beta_{k} + \gamma_{jk}) \\
&= P(Z - \gamma_{jk} < \bm{x}_{ijk}^{T}\boldsymbol\beta_{k}) \\
&= P\Big(\frac{Z - \gamma_{jk}}{\sqrt{1 + \sigma_{k}^2}} < \frac{\bm{x}_{ijk}^{T}\boldsymbol\beta_{k}}{\sqrt{1 + \sigma_{k}^2}}\Big) \\
&= \Phi\Big(\frac{\bm{x}_{ijk}^{T}\boldsymbol\beta_{k}}{\sqrt{1 + \sigma_{k}^2}}\Big)\\
&= \Phi(\bm{x}_{ijk}^{T}\tilde{\bm{\beta}}_{k}),
\end{align}
\noindent where $\tilde{\bm{\beta}}_{k} = \frac{\bm{\beta_{k}}}{\sqrt{1 + \sigma_{k}^2}}$.
\end{document}